\documentclass{article}

\usepackage{authblk}
\usepackage{graphicx}
\usepackage{array}
\usepackage{amssymb}
\usepackage{amsmath}
\usepackage{amsfonts}
\usepackage[all]{xy}
\usepackage{subfig}
\usepackage{fixltx2e}
\usepackage{verbatim}
\usepackage{fancyvrb}
\usepackage{placeins}

\PassOptionsToPackage{hyphens}{url}\usepackage[pdftex,                
bookmarks=true,
bookmarksopen=true,
bookmarksopenlevel=0,
bookmarksnumbered=true,
unicode=false,
hypertexnames=true,
breaklinks=true,
colorlinks=true,%
citecolor=black,%
urlcolor=black,%
menucolor=black,%
linkcolor=black,%
]{hyperref}

\newenvironment{mcrl2}%
{\begin{trivlist}
\item\begin{tabular}{@{}>{\bf}p{2em}L@{\ }L@{\ }L@{\ }L@{\ }L@{\ }L@{\ }L@{\ }L}}%
{\end{tabular}\end{trivlist}}

\newcolumntype{L}{>{$}l<{$}}
 	\newcolumntype{C}{>{$}c<{$}}
 	\newcolumntype{R}{>{$}r<{$}}

\newcommand{\nat}{\mathbb{N}}
\newcommand{\bool}{{\mathbb B}}
\newcommand{\cons}{\ensuremath{\hspace{0.12em}\triangleright\hspace{0.08em}}}

\newcommand{\ap}{{:}}
\newcommand{\List}{\it List}

\newcommand{\xr}{\textit{xr}}
\newcommand{\pl}{\textit{pl}}
\newcommand{\state}{\textit{PSt}}
\newcommand{\eqstate}{\textit{St}}
\newcommand{\xray}{\textit{XRay}}
\newcommand{\plane}{\textit{Plane}}
\newcommand{\din}{\textit{In}}
\newcommand{\dout}{\textit{Out}}
\newcommand{\condfun}{\textit{cond}}
\newcommand{\eval}{\textit{Eval}}
\newcommand{\condthenfun}{\textit{condthen}}
\newcommand{\sq}{\textquotesingle}
\usepackage{textcomp}
\newcommand{\true}{\textit{true}}
\newcommand{\false}{\textit{false}}

\begin{document}

\title{Industrial Experiences with a Formal DSL Semantics to Check Correctness of DSL Transformations}
\date{}

\author[1]{Sarmen Keshishzadeh}
\author[2]{Arjan J. Mooij}
\author[2,3]{Jozef Hooman}
\affil[1]{Eindhoven University of Technology, Eindhoven, The Netherlands}
\affil[2]{Embedded Systems Innovation by TNO, Eindhoven, The Netherlands}
\affil[3]{Radboud University Nijmegen, Nijmegen, The Netherlands}

\maketitle

\begin{abstract}
A domain specific language (DSL) abstracts from implementation details and is aligned with the way domain experts reason about a software component. The development of DSLs is usually centered around a grammar and transformations that generate implementation code or analysis models. The semantics of the language is often defined implicitly and in terms of a transformation to implementation code. In the presence of  multiple transformations from the DSL, the consistency of the generated artifacts with respect to the semantics of the DSL is a relevant issue. We  show that a formal semantics is essential for checking the consistency between the generated artifacts. We exploit the formal semantics in an industrial project and use formal techniques based on equivalence checking and model-based testing for consistency checking. We report  about our experience with this approach in an industrial development project.
\end{abstract}

\section{Introduction}\label{sec:acq_introuduction}
A domain specific language (DSL) \cite{DKV00} abstracts from implementation details and is aligned with the way domain experts reason about a software component. By focusing on the essential concepts in a problem domain, DSLs facilitate the involvement of domain experts in the development of DSL specifications.

Tool support for the development of DSLs is improving constantly. Language workbenches such as Xtext \cite{Xtext} enable language designers to define their languages and develop transformations that generate various artifacts from domain specific models. This has boosted the popularity of DSL approaches in industry; see \cite{MHA13,VLHW13}.

The development of DSLs is usually centered around a grammar and transformations that generate implementation code or analysis models from DSL specifications. In such a setting, the main focus is on the transformation to implementation code which is very valuable in industrial practice. The generated analysis models allow to reason about the behavior described in DSL models. For example, properties can be verified against verification models or simulation models can be used to explore the modeled behavior interactively.  

In DSL approaches, the semantics of the language is  often defined implicitly and in terms of the generated implementation. Although DSLs focus on the essential concepts of their respective domains, the semantics of the language constructs are not always obvious. The lack of a formal semantics can give rise to different interpretations and this can potentially cause inconsistencies between the transformations.

Various authors \cite{ABE11,BFLM05} have proposed to use a formal semantics to describe the precise meaning of the language constructs. The formalization also allows them
to have a single reference that should be followed by all the transformations.



Implementing a transformation from a DSL to a programming/modeling language is not a trivial task. Having a formal semantics as a reference does not guarantee consistency between multiple transformations from a DSL. The developer of a transformation should have a deep understanding of the DSL and the target language and construct a transformation that does not deviate from the semantics of the DSL. In \cite{BG13} the authors indicate that such tasks are very error prone and introducing redundancy is an effective way to reduce the rate of mistakes. 

We propose to use the formal semantics of a DSL and introduce redundancy using the following formal techniques:
\begin{itemize}
  \item equivalence checking (by transforming models to a formalism that allows checking equivalence of behaviors);
  \item model-based testing (by testing conformance of executable models to a test model).
\end{itemize}

In this paper, we report on our experiences with using these redundant mechanisms in the context of an industrial DSL. The results obtained in the project show that these formal techniques can effectively detect inconsistencies between transformations from a DSL. To hide the complexities of these techniques, we have  developed a push-button technology that allows industrial users to automatically perform these checks for the artifacts generated from a specification in the DSL. 

Some authors \cite{EE08} propose to formally prove the correctness of transformations. Since in practice transformations are improved regularly, we use automated techniques based on equivalence checking and model-based testing to validate the consistency between the artifacts, as opposed to formally validating the transformations themselves.

\paragraph{Overview}
We discuss preliminaries in Section~\ref{sec:acq_pre}. In Section~\ref{sec:acq_casestudy} we describe an industrial control component and informally introduce a DSL for describing its behavior; the abstract syntax and semantics of the language are discussed in Section~\ref{sec:acq_semantics}. In Section~\ref{sec:acq_poosl} we consider the formalized semantics and construct a transformation from DSL models to models that enable simulation. In Section~\ref{sec:acq_mbt} we generate test models from DSL specifications and apply model-based testing to assess the quality of an implementation of the industrial component.  The correctness of the used models with respect to the semantics is validated in Section~\ref{sec:acq_comparison}. In Section~\ref{sec:acq_transformations} we discuss about our experiences with different types of model transformations. In Section~\ref{sec:acq_related} we discuss related work. Section~\ref{sec:acq_conclusion} contains conclusions and  directions for future research. 

\section{Preliminaries}\label{sec:acq_pre}
In this section we give an overview of the micro Common Representation Language 2 (mCRL2) \cite{GM14}. mCRL2 is a process algebra that extends the Algebra of Communicating Processes (ACP) \cite{BK84} with data and time. The mCRL2 language and its supporting toolset \cite{mCRL2} can be used to specify and analyze the behavior of distributed systems and protocols.

In this short overview, we focus on the language constructs that we need throughout this paper. The interested reader can refer to \cite{GM14} for a detailed explanation about this process algebra. We first explain the way data types are defined and used in mCRL2 (Section~\ref{subsec:acq_dt}). Then, we describe behavioral specifications in the language (Section~\ref{subsec:acq_proc_spec}).

\subsection{Data Specification}\label{subsec:acq_dt}
mCRL2 offers ways to specify data types (also known as sorts) and use their elements in specifications. Standard data types such as natural numbers ($\nat$) and booleans ($\bool$) are predefined in the language. Common operations on these data types are also available (e.g., $\approx$ denotes equality  for $\nat$ and $\bool$).  

Besides the standard data types, the user can define new data types in a specification. One way to declare a new data type is to explicitly characterize its elements in a structured data type. For instance, we can define a structured sort called ${\it Color}$ with the elements ${\it Red}$, ${\it Green}$, and ${\it Blue}$ as follows:
\begin{mcrl2}
sort & {\it Color} ~ = ~ {\bf struct}~~{\it Red}~|~{\it Green}~|~{\it Blue};
\end{mcrl2}
It is also possible to declare structured types that depend on other data types. 
For instance, a data type called ${\it Message}$ that contains pairs of natural numbers can be defined as follows:
\begin{mcrl2}
sort & {\it Message} ~ = ~ {\bf struct}~~{\it Pair}({\it fst}\ap \nat,{\it snd}\ap \nat);
\end{mcrl2}
Elements of ${\it Message}$ have the shape ${\it Pair}(n_1,n_2)$ where $n_1,n_2\in \nat$. The declaration of ${\it Message}$ provides two projection functions, namely, ${\it fst}$ and ${\it snd}$. These functions can extract the elements of ${\it Pair}(n_1,n_2)$:
\begin{align*}
{\it fst}({\it Pair}(n_1,n_2))=n_1 ~~~~~~~~~~ {\it snd}({\it Pair}(n_1,n_2))=n_2
\end{align*}

Functions can also be declared and used in the mCRL2 language. Given two data types $A$ and $B$, the notation $A\rightarrow B$ denotes the sort of functions from $A$ to $B$. The mCRL2 language includes a specific operator called function update for unary functions. For a function $f\in A\rightarrow B$ this operation is denoted by $f[a\rightarrow b]$. The notation represents a function that maps $a$ to $b$ and maps all the other elements of $A$ like $f$ does. 

We can declare the sort of functions from $\nat$ to $\nat$ as follows:  
\begin{mcrl2}
sort & {\it NatFunc} ~ = ~ \nat \rightarrow \nat;
\end{mcrl2}
We consider two examples of this sort:
\begin{itemize}
\item
${\it succ}$: given $n\in \nat$ returns $n+1$;
\item
${\it condsqr}$: given $n\in \nat$ returns $n^2$ when $n>10$; otherwise, it returns $n$.
\end{itemize}
We declare these functions using the ${\bf map}$ keyword:
\begin{mcrl2}
map  & {\it succ}, {\it condsqr}~\ap~{\it NatFunc};
\end{mcrl2}
To define ${\it succ}$ and ${\it condsqr}$, it is necessary to specify how calculations are performed in these functions. This is realized by introducing equations using the keyword ${\bf eqn}$. Variables that are used in the equations are declared by the keyword ${\bf var}$. 
\begin{mcrl2}     
var  & n~\ap~\nat;\\
eqn  & {\it succ}(n)=n+1;\\
     & {\it condsqr}(n)={\it if}(n>10,n^2,n);
\end{mcrl2}
The conditional operation has the shape ${\it if}(c,t,u)$. It evaluates to the term $t$ if the condition $c$ holds and it evaluates to the term $u$ if $c$ does not hold. 


Lists are also a built-in data-type in mCRL2. The set of lists where all elements are from a sort $A$ are represented by $\List(A)$. Elements of $\List(A)$ are built with two constructors: $[]$ the empty list and $a\cons \ell$ which puts $a$ (of type $A$) in front of list $\ell$ (of type $\List(A)$). A list can be defined explicitly by specifying its elements and putting them between square brackets. For example, $[22,4]$ is a list of natural numbers. 

\subsection{Process Specification}\label{subsec:acq_proc_spec}
The mCRL2 language allows us to specify behavior using a small set of primitives and operators. We use a simple example to describe some basic constructs that can be used to specify behavior in mCRL2. The example is a modulo $3$ counter that starts counting from $0$ and resets itself when it reaches $3$. 

In mCRL2, behavior is described in terms of processes. Actions are elementary processes and they represent observable atomic events. Actions can carry data parameters. In our simple example, ${\it count}$ and ${\it reset}$ can be considered as actions performed by the counter. For example, ${\it count}$ carries a data parameter to indicate the current number. These actions are declared as follows:
\begin{mcrl2}
act & {\it count}~\ap~\nat;\\
    & {\it reset};
\end{mcrl2}

Actions can be combined using different operators to form processes that specify more complex behavior. For instance, the non-deterministic choice between process $p$ and $q$ is denoted by $p+q$. The sequential composition of $p$ and $q$ is denoted by $p.q$. The resulting process first performs the behavior of $p$ and then behaves as $q$. Data values can also influence the course of actions. Suppose $c$ is a boolean expression. The process $c\rightarrow p\diamond q$ behaves as $p$ if $c$ holds and otherwise it behaves as $q$. The ``else'' part of the conditional operator can be omitted. If $c$ does not hold in $c\rightarrow p$, deadlock will occur. 

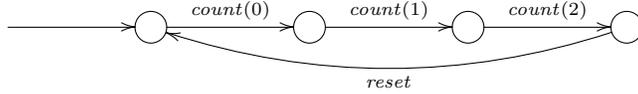
\begin{figure}
\centerline{
\xymatrix@C=4pc{
\ar[r] & *++[o][F]{} \ar[r]^{{\it count}(0)} & *++[o][F]{} \ar[r]^{{\it count}(1)} & *++[o][F]{} \ar[r]^{{\it count}(2)} & *++[o][F]{} \ar@/^1.3pc/[lll]^{{\it reset}}
}
}
\caption{Behavior of ${\it Counter}$}
\label{fig:acq_lts}
\end{figure}

The following process expression specifies the modulo $3$ counter. The process ${\it Counter}$ carries a data parameter to keep track of the current number. The initial behavior is specified by ${\bf init}$, i.e., counting starts from $0$.
\begin{mcrl2}
proc & {\it Counter}(n\ap\nat)=(n<3)\rightarrow {\it count}(n).{\it Counter}(n+1)  \\
     &\phantom{{\it Counter}(n\ap\nat)}+(n \approx 3)\rightarrow {\it reset}.{\it Counter}(0);\\
init & {\it Counter}(0);     
\end{mcrl2}
For any $n\leq 3$ exactly one of the conditions $n<3$ and $n\approx 3$ will evaluate to ${\it true}$. The process performs action ${\it count}(n)$ when $n<3$ and then behaves as ${\it Counter}(n+1)$. When $n\approx 3$, the process performs action ${\it reset}$ and starts counting from $0$. Fig.~\ref{fig:acq_lts} depicts the labeled transition system of the counter.

\section{Application: a Clinical X-ray Generator}\label{sec:acq_casestudy}

In this section we introduce an industrial control component that we use for reporting our experiences (Section~\ref{subsec:acq_ph}). We also informally describe a DSL for specifying the behavior of the component (Section~\ref{subsec:acq_dsl}).

\subsection{Platform}\label{subsec:acq_ph}

\begin{figure}[t]
\centering
\subfloat[][Interventional X-ray]
{\includegraphics[height=4.2cm]{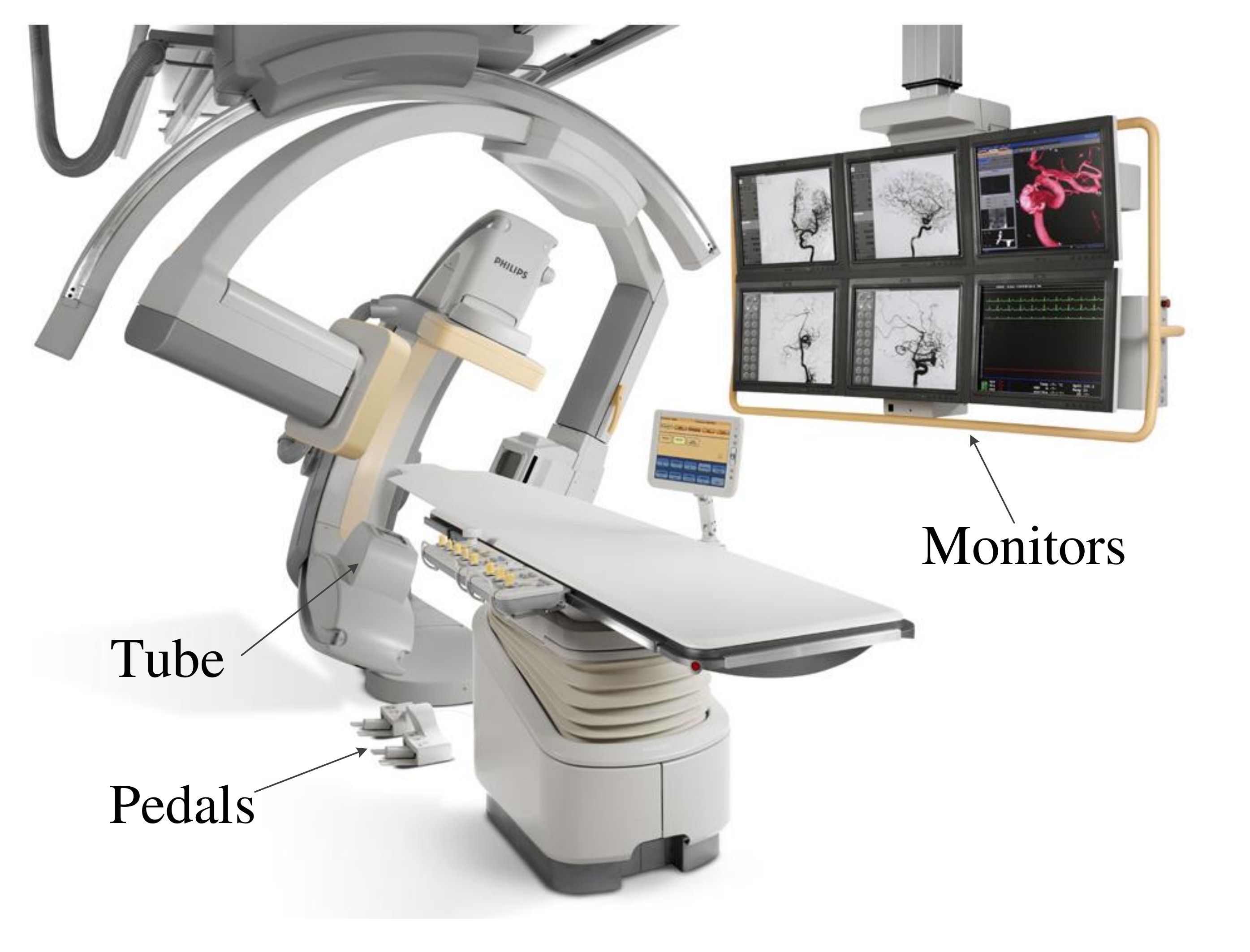}\label{fig:acq_ixr}}
\qquad
\subfloat[][Interfaces of \textit{Pedal Handling}]{\includegraphics[height=4.2cm]{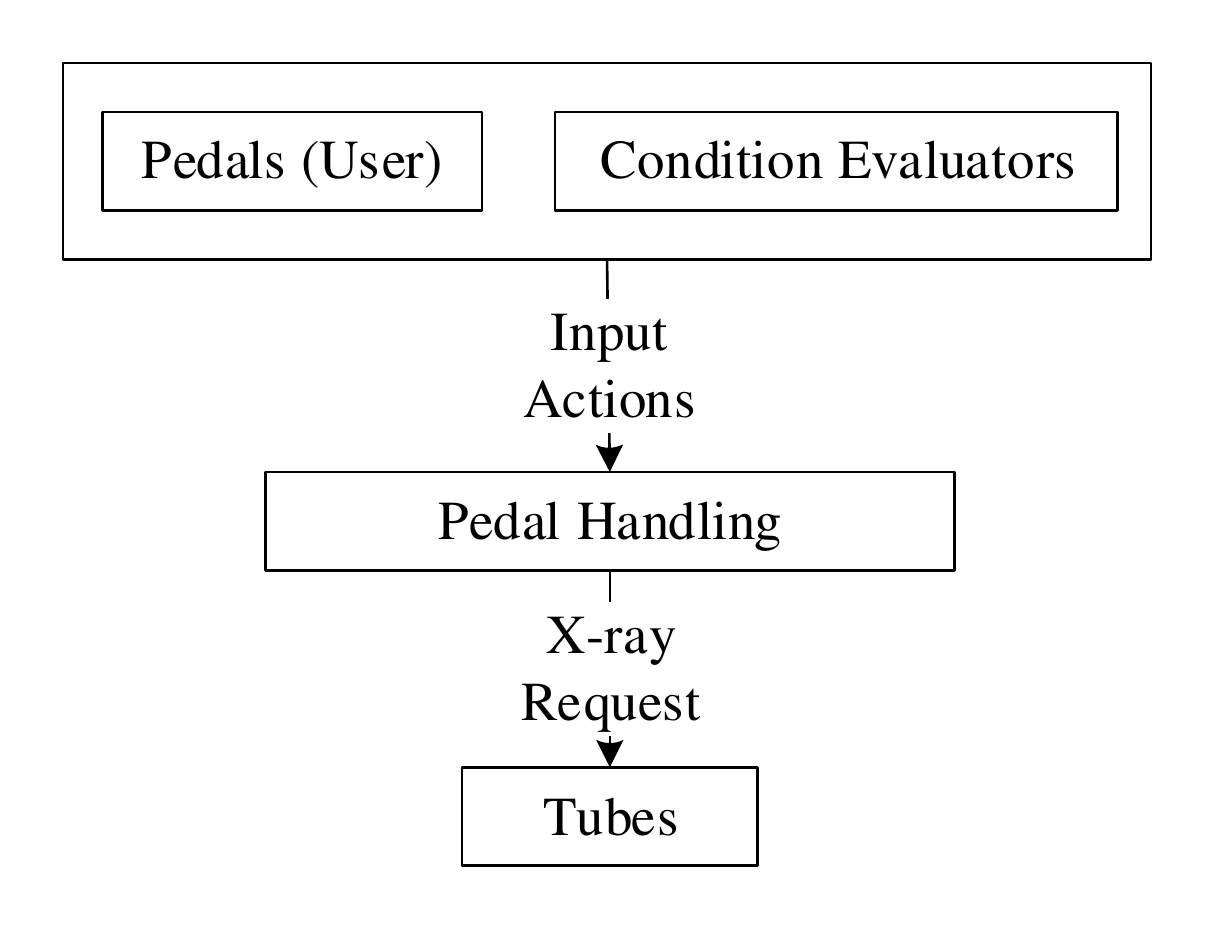}\label{fig:acq_layers}}
\caption{Industrial Study Case}
\end{figure}

Philips Healthcare produces interventional X-ray systems (see Fig.~\ref{fig:acq_ixr}) which are used to perform minimally-invasive medical procedures. During a procedure, the surgeon uses the images on the monitors as guidance. Images are constructed for two projections. The X-ray system of Fig.~\ref{fig:acq_ixr} consists of two planes: frontal (top-down) and lateral (left-right). These planes can be used separately or together (biplane).

The surgeon can send X-ray requests using the pedals. The interventional system of Fig.~\ref{fig:acq_ixr} includes a component called \textit{Pedal Handling} that makes decisions about the amount of X-ray that should be generated in the tube of each plane (Fig.~\ref{fig:acq_layers}). From each plane the following types of X-ray can be generated:
\begin{itemize}
\item
Fluoroscopy: low dose X-ray, for obtaining real-time images;
\item
SingleShot: high dose X-ray, for capturing a single image;
\item
Series: high dose X-ray, for recording a series of images.
\end{itemize}

\textit{Pedal Handling} also takes into account conditions that should interrupt the X-ray, or that should prevent the X-ray from starting. \textit{Condition Evaluators} continuously evaluate these conditions and notify \textit{Pedal Handling} when changes occur.

\subsection{DSL for Pedal Handling}\label{subsec:acq_dsl}

To describe the behavior of \textit{Pedal Handling}, domain experts mainly focus on the external interfaces of the component. Starting from the initial state, they think about input actions received from \textit{Pedals} and \textit{Condition Evaluators}. Based on the received input, the component might change its current state; it also makes a decision about the output X-ray and sends a request to the tubes. This process continues by receiving the next input. Thus, from the domain expert's point of view the behavior of \textit{Pedal Handling} can be described with alternating sequences of input and output actions. 

To specify the behavior of \textit{Pedal Handling}, we use a DSL that fits this way of reasoning. We have used Xtext \cite{Xtext} to define the grammar (concrete syntax) of this DSL. Based on the grammar a parser, an Eclipse-based editor, and a set of Java classes describing the concepts in the abstract syntax of the language are automatically generated.  

\begin{figure}[t]
\centering
\includegraphics[scale=0.6]{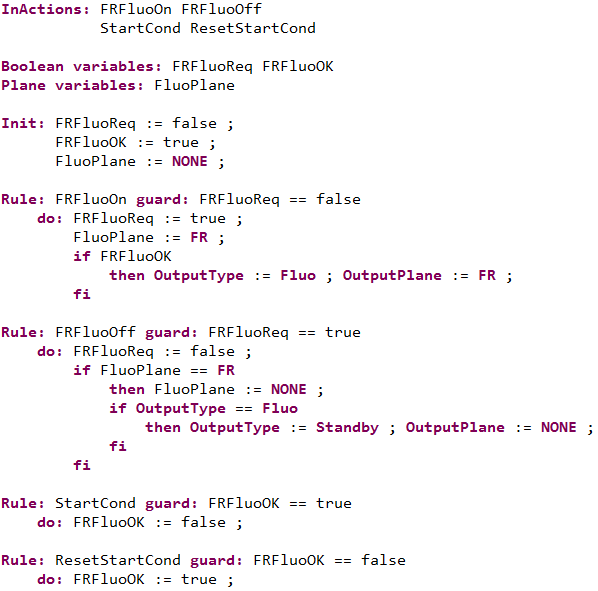}
\caption{Snapshot of a DSL Model\label{fig:acq_dslexample}}
\end{figure}

Fig.~\ref{fig:acq_dslexample} depicts a simplified specification in the DSL\footnote{For confidentiality reasons, details have been changed in this example.}. A DSL model starts with declaring the input actions that can be received by \textit{Pedal Handling} (\texttt{InActions}). Afterwards, it declares variables that keep track of the current state of the component (\texttt{Boolean variables} and \texttt{Plane variables}) and their initial values (\texttt{Init}). Two not-explicitly-declared variables, i.e., \texttt{OutputType} and \texttt{OutputPlane}, determine the output of \textit{Pedal Handling}. 

The internal logic of the component is described in terms of \texttt{Rule}s. Each rule refers to an input action and consists of a guard and a do clause. The guard describes when the input action is enabled and the do clause determines how the action influences the state of the component and the output. A DSL model specifies exactly one rule for each input action. Multiple rules for an action are not supported.


Rules of the DSL use very simple constructs for describing behavior. However, the precise meaning of some constructs is not immediately obvious. For instance, it is not obvious whether evaluating the do clause of an action can be interrupted by receiving a new input action. Since a do clause may contain multiple assignments to a variable, it is also relevant to determine when a variable assignment takes effect.

In Section~\ref{sec:acq_semantics} we identify the main concepts in the DSL (abstract syntax) and give a formalization which clearly specifies the semantics of the language. For instance, our semantics enforces that do clauses should be interpreted in an atomic way and each assignment to a variable immediately changes its value such that the previous value is overwritten.

\section{Abstract Syntax and Semantics of the DSL}\label{sec:acq_semantics}
In Section~\ref{subsec:acq_dsl} we informally introduced a DSL and motivated the need for a formal semantics that describes the meaning of the language constructs. In this section, we identify the main concepts of the language and give a precise meaning to them by means of a transformation to a language, namely mCRL2, that has a formal semantics. The semantics of the mCRL2 language is defined in \cite{GM14} using Structural Operational Semantics (SOS) \cite{P04}.

We use the model from Fig.~\ref{fig:acq_dslexample} as a running example to describe our transformation to mCRL2. Our choice of mCRL2 is  motivated by the expressiveness of the language, the availability of a toolset \cite{mCRL2} that supports analysis of behavior, and our previous experience \cite{KGK13} with the language and toolset.

We first discuss the required data specifications in Section~\ref{subsec:acq_domain}. Then we use process expressions to describe the behavior specified by DSL models in Section~\ref{subsec:acq_semantics}. In Section~\ref{subsec:acq_domain} and \ref{subsec:acq_semantics}, we have dedicated separate paragraphs to discuss the concepts from the abstract syntax of the DSL. Bold text at the start of each paragraph indicates the name of the concepts. 

Finally, in Section~\ref{subsec:acq_analysis} we discuss a few correctness properties that we verify against DSL models.

\subsection{Data Specification}\label{subsec:acq_domain}
\paragraph{Plane, X-Ray Type} As mentioned in Section~\ref{subsec:acq_ph}, X-ray can be generated from three planes (frontal, lateral, and biplane). We define the data types ${\it Plane}$ and ${\it XRay}$ to describe the planes and the type of X-ray generated from them.  
\begin{mcrl2}
sort & {\it Plane} ~ = ~{\bf struct}~~{\it None} ~|~ {\it FR}~|~{\it LT}~|~{\it BI};\\
     & {\it XRay} ~ = ~{\bf struct}~~{\it Standby} ~|~{\it Fluo}~|~{\it SingleShot}~|~{\it Series};
\end{mcrl2}
The combination of ${\it None}$ and ${\it Standby}$ describes a situation in which no X-ray is generated from the planes. 

\paragraph{State} A DSL model declares a set of boolean and plane variables. The valuation of these variables and the two special variables \texttt{OutputType} and \texttt{OutputPlane} determine the state of the transition system described by the DSL model. 

To describe the notion of state in mCRL2, we create two structured data types ${\it B}$ (for boolean variables) and ${\it P}$ (for plane variables). The structure of these sorts corresponds to the variable declarations of the DSL model. For the model of Fig.~\ref{fig:acq_dslexample}, $B$ and $P$ are defined as follows:
\begin{mcrl2}
sort & {\it B}~=~{\bf struct}~~{\it FRFluoReq}~|~{\it FRFluoOK};\\
     & {\it P}~=~{\bf struct}~~{\it FluoPlane};
\end{mcrl2}
To specify the valuations of boolean and plane variables, we declare the following data types:
\begin{mcrl2}
sort & {\it BVals} ~=~ {\it B}\rightarrow \bool;\\
& {\it PVals} ~=~ {\it P}\rightarrow {\it Plane};
\end{mcrl2}
Finally, the notion of state can be formalized as follows:
\begin{mcrl2}
sort & {\it PSt}~=~{\bf struct}~~{\it St}(\mathit{bs}\ap\mathit{BVals}, {\it ps}\ap{\it PVals}, {\it outType}\ap {\it XRay}, {\it outPlane}\ap {\it Plane});
\end{mcrl2}
The projection functions ${\it outType}$ and ${\it outPlane}$ extract the values of \texttt{OutputType} and \texttt{OutputPlane} from states.

In a DSL model, the initial values of the boolean and plane variables are specified by \texttt{Init}. We declare $bs_0\in {\it BVals}$ and $ps_0\in {\it PVals}$ to specify the initial values in mCRL2. We also declare $s_0\in \state$ to describe the initial state.   
\begin{mcrl2}
map & bs_0: {\it BVals};\\
    & ps_0: {\it PVals};\\
    & s_0: \state;\\
eqn & bs_0({\it FRFluoReq})={\it false};\\
    & bs_0({\it FRFluoOK})={\it true};\\
    & ps_0({\it FluoPlane})={\it None};\\
    & s_0={\it St}(bs_0,ps_0,{\it Standby},{\it None});    
\end{mcrl2}
The initial values of \texttt{OutputType} and \texttt{OutputPlane} cannot be specified by \texttt{Init} in the DSL. It is assumed that initially no X-ray is generated from the planes. Hence, we specify $s_0$ such that:
\begin{align*}
{\it outType}(s_0)={\it Standby} ~~~~~~~~~~ {\it outPlane}(s_0)={\it None}
\end{align*}

\paragraph{Guard, Do Clause, Rule} A DSL model specifies one rule for each input action. The rule of an input action consists of a guard and a do clause. From Fig.~\ref{fig:acq_dslexample} one can see that a guard is a function from states to booleans. A do clause consists of a sequence of assignments/conditionals that given the current state can change the values of the variables and produce a new state. We describe guards, do clauses, and rules as follows:
\begin{mcrl2}
sort & {\it Guard} ~ = ~ {\it PSt} \rightarrow \bool;\\
     & {\it DCl} ~ = ~ \List({\it PSt} \rightarrow {\it PSt});\\
     & {\it Rule} ~ = ~{\bf struct}~~ R({\it gr}\ap {\it Guard}, {\it dc}\ap{\it DCl});
\end{mcrl2}
The projection functions ${\it gr}$ and ${\it dc}$ extract the guard and do clause element of a rule, respectively. 

For the DSL model of Fig.~\ref{fig:acq_dslexample} with $4$ rules, we declare the guards, do clauses, and rules $g_i,d_i,r_i$ for $1\leq i\leq 4$: 
\begin{mcrl2}
map & g_1,g_2,g_3,g_4~\ap~ {\it Guard};\\
    & d_1,d_2,d_3,d_4~\ap~ {\it DCl};\\
    & r_1,r_2,r_3,r_4~\ap~ {\it Rule};
\end{mcrl2}
The calculations of each guard can be described in terms of an equation. For instance, the guard of the first rule of Fig.~\ref{fig:acq_dslexample} can be defined as follows:
\begin{mcrl2}
var & b:{\it BVals};\\
    & p:{\it PVals};\\
    & \xr:{\it XRay};\\
    & \pl:{\it Plane};\\
eqn & g_1(\eqstate(b,p,\xr,\pl))=(b({\it FRFluoReq})\approx {\it false});
\end{mcrl2}

To describe do clauses, we   specify assignments and conditionals in terms of equations. To explain this, we consider the do clause of the first rule from Fig.~\ref{fig:acq_dslexample}. This do clause contains four assignments and one conditional. 

We describe each assignment as a function that updates one of the components of a state $\eqstate(b,p,\xr,\pl)$. We denote the assignments of the first rule by $a_1,a_2,a_3,a_4$ based on their order of appearance:
\begin{mcrl2}
map & a_1,a_2,a_3,a_4~\ap ~ \state\rightarrow \state;\\   
eqn & a_1(\eqstate(b,p,\xr,\pl))=\eqstate(b[{\it FRFluoReq}\rightarrow \true],p,\xr,\pl);  \\
    & a_2(\eqstate(b,p,\xr,\pl))=\eqstate(b,p[{\it FluoPlane \rightarrow {\it FR}}],\xr,\pl);  \\
    & a_3(\eqstate(b,p,\xr,\pl))=\eqstate(b,p,{\it Fluo},\pl); \\
    & a_4(\eqstate(b,p,\xr,\pl))=\eqstate(b,p,\xr,{\it FR});
\end{mcrl2}
For example, the first assignment updates the values of boolean variables by setting ${\it FRFluoReq}$ to ${\it true}$. 

A conditional statement of a do clause is specified by a term of the shape ${\it if}(c,t,u)$ in equations. The conditional in the first rule of Fig.~\ref{fig:acq_dslexample} can be described as follows:
\begin{mcrl2}
map & \condfun ~\ap~\state \rightarrow \state;\\   
eqn & \condfun (\eqstate(b,p,\xr,\pl))= {\it if}(b({\it FRFluoOK}),\eval(\condthenfun,\eqstate(b,p,\xr,\pl))\\
    &\phantom{cond(\eqstate(b,p,\xr,\pl))= {\it if}(b({\it FRFluoOK}},\eqstate(b,p,\xr,\pl));
\end{mcrl2}
In this equation, $\eval$ is a function that evaluates a sequence of assignments or conditionals and $\condthenfun$ is the ``then'' part of the conditional (see below).

A do clause is described as a sequence of assignments/conditionals. For example, we describe the do clause of the first rule ($d_1$) as a sequence of the assignments $a_1,a_2$ and the conditional $\condfun$. The ``then'' part of a conditional is also specified as a sequence of its components. The ``then'' part of the conditional in the first rule ($\condthenfun$) is a sequence of $a_3,a_4$: 
\begin{mcrl2}
map & \condthenfun~\ap~\List(\state\rightarrow\state);\\
eqn & d_1=[a_1,a_2,\condfun];\\
    & \condthenfun=[a_3,a_4];
\end{mcrl2}

Having sequences of assignments and conditionals, it is also essential to define a function that applies a sequence of statements to a state and returns the resulting state. The following mCRL2 description defines ${\it Eval}$ for this purpose: 
\begin{mcrl2}
map & \eval~\ap~ \List (\state\rightarrow \state)\times\state \rightarrow \state;\\
var & s~\ap~\state;\\
    & f~\ap~\state\rightarrow\state;\\
    & \ell~\ap~\List(\state\rightarrow\state);\\
eqn & \eval([],s)=s;\\
    & \eval(f\cons \ell,s)=\eval(\ell,f(s));
\end{mcrl2}
The function $\eval$ is defined by specifying its effect on terms of the shape $[]$ and $f\cons \ell$ (the constructors of $\List$).

Finally, each rule consists of a guard and a do clause. For example, the first rule from Fig.~\ref{fig:acq_dslexample} (declared as $r_1$ above) can be defined as follows:
\begin{mcrl2}
eqn & r_1=R(g_1,d_1);
\end{mcrl2}

\subsection{Process Specification}\label{subsec:acq_semantics}
\paragraph{Action} The \textit{Pedal Handling} component performs two types of actions: input and output. A DSL model explicitly declares a set of input actions (from \textit{Pedals} and \textit{Condition Evaluators}) by \texttt{InActions}. \textit{Pedal Handling} also performs actions $\textit{output}(\xr,p)$ to send requests for $\xr\in \xray$ to $p\in \plane$. This action is not explicitly declared in the DSL. Corresponding to the input actions and the output action we declare actions in mCRL2. For Fig.~\ref{fig:acq_dslexample} we declare:
\begin{mcrl2}
act & {\it FRFluoOn}, {\it FRFluoOff}, {\it StartCond}, {\it ResetStartCond};\\
    & {\it output}~\ap ~{\it XRay}\times {\it Plane};
\end{mcrl2}

In Section~\ref{subsec:acq_dsl}, we mentioned that domain experts  describe the behavior of \textit{Pedal Handling} with alternating sequences of input and output actions. The semantics of the DSL is aligned with this intuition. We specify the semantics of the DSL model of Fig.~\ref{fig:acq_dslexample} in terms of the following processes:
\begin{mcrl2}
proc &~ {\it P_{In}}(s\ap {\it PSt})= {\it gr}(r_1)(s) \rightarrow ({\it FRFluoOn}.{\it P_{\it Out}}({\it Eval}({\it dc}(r_1),s)))  \\
     &\phantom{{\it P_{In}}(s\ap {\it PSt})~}+ {\it gr}(r_2)(s) \rightarrow ({\it FRFluoOff}.{\it P_{\it Out}}({\it Eval}({\it dc}(r_2),s)))\\
	&\phantom{{\it P_{In}}(s\ap {\it PSt})~}+ {\it gr}(r_3)(s) \rightarrow ({\it StartCond}.{\it P_{\it Out}}({\it Eval}({\it dc}(r_3),s)))\\     
	&\phantom{{\it P_{In}}(s\ap {\it PSt})~}+ {\it gr}(r_4)(s) \rightarrow ({\it ResetStartCond}.{\it P_{\it Out}}({\it Eval}({\it dc}(r_4),s)));\\     
& {\it P_{Out}}(s\ap {\it PSt}) = {\it output}({\it outType}(s), {\it outPlane}(s)).P_{\it In}(s);\\
init & {\it P_{In}}(s_0);     
\end{mcrl2}

The process $P_{\din}$ describes the behavior of \textit{Pedal Handling} when the component is ready to receive an input. It carries a data parameter that indicates the current state. The process $P_{\din}$ uses a combination of choices and conditional operators for case distinction. The guards (extracted by ${\it gr}(r_i)$) are used as conditions in the conditional operators to determine enabled actions. Performing an input action updates the state based on the corresponding do clause (extracted by ${\it dc}(r_i)$). 

The process $P_{\dout}$ describes the behavior of \textit{Pedal Handling} when the component is ready to produce an output. In this situation, ${\it output}$ is performed and it carries the X-ray type and plane extracted from the current state. Performing the output action does not influence the state.

The processes $P_{\din}$ and $P_{\dout}$ enforce alternating execution of the input and output actions. Do clauses are evaluated by $\eval$. Thus, new actions cannot be performed before a do clause is completely evaluated. Moreover, assignments have an immediate effect.

\subsection{Validating DSL Models}\label{subsec:acq_analysis}
For a safety critical component, it is desired to use DSL models as a single source to automatically obtain models that enable analysis using various formal techniques. 

To facilitate formal analysis and verification, we have automated the transformation from the DSL to mCRL2. We have  used the mCRL2 toolset to generate the state spaces of DSL models and to verify properties expressed in a variant of the modal $\mu$-calculus \cite{K83}. An actual DSL model declares $25$ input actions and their effects in terms of rules. The corresponding state space consists of approximately $45000$ states and $350000$ transitions. We have verified some safety properties against this model, e.g., ``deadlock-freedom'', ``no X-ray is generated from the planes when there is no request from the user''. A formalization of some safety properties for the DSL model of Fig.~\ref{fig:acq_dslexample} are included in Appendix~\ref{app:properties}. 

\begin{figure}[t]
\centering
\includegraphics[scale=0.53]{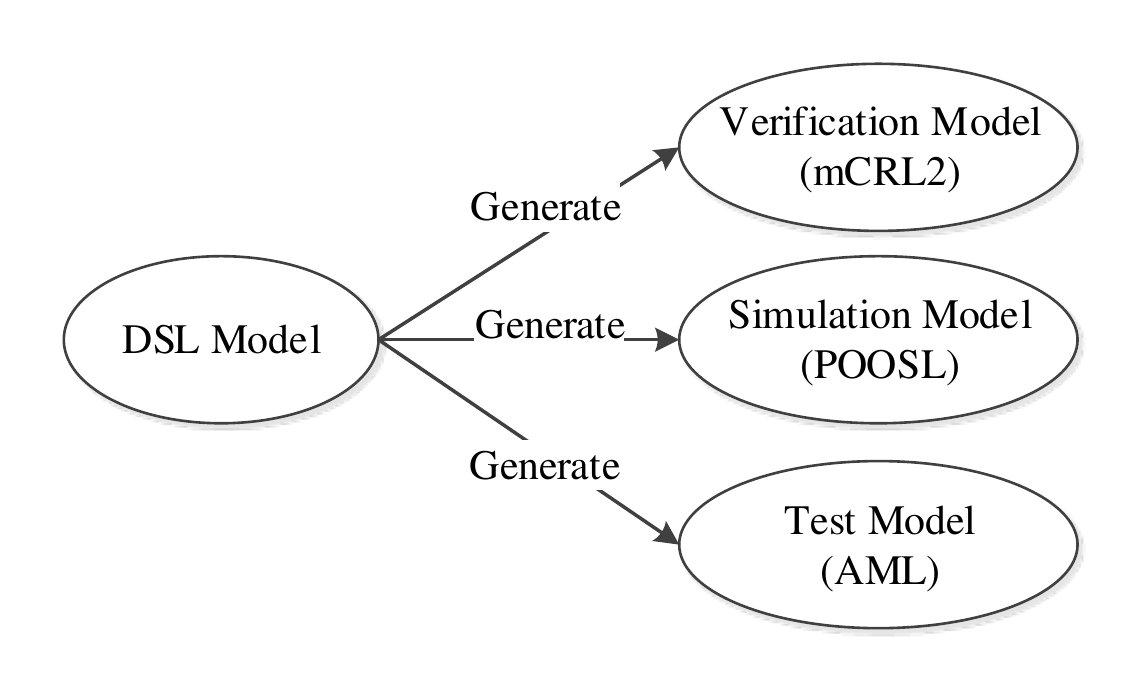}
\caption{Generated Artifacts for \textit{Pedal Handling}\label{fig:acq_artifacts}}
\end{figure}

The mCRL2 formalization of the semantics can also be used as a reference to implement transformations to formalisms that support other kinds of analysis. In Section~\ref{sec:acq_poosl} and \ref{sec:acq_mbt} we discuss automated transformations from the DSL to formalisms that enable simulation and model-based testing, respectively. Fig.~\ref{fig:acq_artifacts} depicts an overview of the transformations from the DSL that we discuss throughout this paper. We have used Xtend \cite{Xtend} to develop these transformations. 

Using the DSL semantics as guidance for implementing such transformations does not give a robust connection between the generated models. However, it helps to avoid obvious inconsistencies between them. In Section~\ref{sec:acq_comparison}, we formally validate the consistency between the generated models.

A proposed implementation for \textit{Pedal Handling} is already available. Thus, at the moment we do not generate implementation code from the DSL.

\section{Obtaining Simulation Models from DSL Models}\label{sec:acq_poosl}
In this section we use the semantics of Section~\ref{sec:acq_semantics} as a reference to construct a transformation from the DSL to models that enable interactive simulation of behavior. We aim to use models in the parallel object-oriented specification language \cite{TFGHPV07} (POOSL) for simulation. Tool support \cite{POOSL} and our previous experience with the language motivated our choice. 

POOSL is a light-weight modeling language and its formal semantics is expressed in terms of timed probabilistic labeled transition systems \cite{B02}. The tools available for POOSL allow us to simulate models in the POOSL language and discuss the modeled behavior with domain experts. Thus, it is desired to have an automated transformation that generates simulation models from DSL specifications. 

In this section we first provide a brief overview of the POOSL language (Section~\ref{subsec:acq_poosl_intro}). Afterwards, we discuss our transformation from the DSL to simulation models in POOSL (Section~\ref{subsec:acq_poosl_trans}).  
\subsection{An Overview of POOSL}\label{subsec:acq_poosl_intro}
In this section we discuss a few constructs of the POOSL language; our focus is on the syntactic constructs that we need for realizing the semantics of Section~\ref{sec:acq_semantics} in POOSL. The interested reader can refer to \cite{B02} and \cite{POOSL} for more details about the language and some instructive examples. We use the modulo $3$ counter from Section~\ref{sec:acq_pre} to illustrate the relevant constructs of the language. 

\begin{figure}[!tp]
\centering
\includegraphics[scale=0.6]{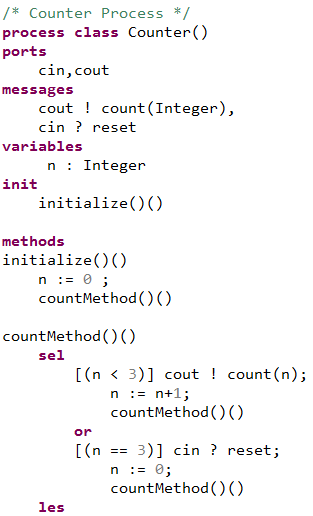}
\caption{A Modulo $3$ Counter in POOSL \label{fig:acq_counter_poosl}}
\end{figure}

In POOSL behavior is described in terms of process classes. Process classes are templates that need to be instantiated to obtain concrete instances that are executable. Each process class declares a set of communication ports that are available to its instances for sending/receiving messages. Process classes also define the signature of messages that can travel across the declared ports. For instance, Fig.~\ref{fig:acq_counter_poosl} depicts a process class called \texttt{Counter} with two ports, namely, \texttt{cin} and \texttt{cout}. The message declarations indicate that messages \texttt{count} and \texttt{reset} can be communicated through the ports \texttt{cout} and \texttt{cin}, respectively. The message \texttt{count} carries an integer parameter to indicate the current number. 

In POOSL there is a notion of initiative for messages, whereas in mCRL2 we do not explicitly define the direction of interactions for actions. The symbols \texttt{!} (message send) and \texttt{?} (message receive) denote the direction of communication for each message. For our simple example, we choose to declare \texttt{count} as a message sent by \texttt{Counter} and \texttt{reset} as a message that can be received by \texttt{Counter}. 

Process classes also consist of variables. Variables of a process are only accessible by the process itself and keep track of data items that are relevant for describing the behavior of the process. Declaring a variable involves specifying a name and its data type. The process \texttt{Counter} in Fig.~\ref{fig:acq_counter_poosl} declares variable \texttt{n} of type integer; this variable is used to store the current number. 

The behavior of a process class is specified in terms of its methods. Each process has an initial method specified by \texttt{init}. Generally speaking, the behavior defined by a method is specified by the following syntax:
\begin{align*}
\texttt{
m}&\texttt{(p\textsubscript{1}:d\textsubscript{p\textsubscript{1}},...,p\textsubscript{k}:d\textsubscript{p\textsubscript{k}})(r\textsubscript{1}:d\textsubscript{r\textsubscript{1}},...,r\textsubscript{m}:d\textsubscript{r\textsubscript{m}}) |l\textsubscript{1}:d\textsubscript{l\textsubscript{1}},...,l\textsubscript{n}:d\textsubscript{l\textsubscript{n}}|
}\\
&~~\texttt{S}
\end{align*}
where \texttt{p\textsubscript{i}} (for $1\leq i\leq k$), \texttt{r\textsubscript{i}} (for $1\leq i\leq m$), and \texttt{l\textsubscript{i}} (for $1\leq i\leq n$) are the input, return, and local parameters of the method, respectively. For each parameter, the corresponding data type is also specified.  

The body of a method (denoted by \texttt{S} in the example above) is composed of different types of statements. In what follows, we discuss a few types of statements that can appear inside a method. Suppose \texttt{x} is a process variable, \texttt{E} is an expression, \texttt{p} is a port, \texttt{S,S\textsubscript{1},S\textsubscript{2}} are statements, and \texttt{C} is a boolean expression. The statement \texttt{x:=E} evaluates the expression \texttt{E} and assigns its value to \texttt{x}. Here we assume that \texttt{E} is a well-defined expression in terms of process variables, constants from standard data types (such as natural numbers and booleans), and common operations on these data types (such as addition for integers); see \cite{B02} for a detailed description of POOSL's data layer and the syntax of expressions. 

The statement \texttt{p!m} denotes a message \texttt{m} being sent to the port \texttt{p}.  In POOSL communication is synchronous, i.e., a send statement \texttt{p!m} is blocked until a matching receive statement \texttt{q?m} can be executed such that the ports \texttt{p} and \texttt{q} are connected via a channel; see description below about connecting ports with channels.

The basic statements mentioned above can be combined in different ways to obtain more complex behaviors. The non-deterministic choice between \texttt{S\textsubscript{1}} and \texttt{S\textsubscript{2}} is denoted by \texttt{sel S\textsubscript{1} or S\textsubscript{2} les}. The notation \texttt{S\textsubscript{1};S\textsubscript{2}} denotes the sequential composition of \texttt{S\textsubscript{1}} and \texttt{S\textsubscript{2}}. 

The guarded command \texttt{[C]S} evaluates the condition \texttt{C} and only if \texttt{C} evaluates to $\true$ the statement \texttt{S} is allowed to execute. Otherwise, deadlock will occur. The statement \texttt{if C then S\textsubscript{1} else S\textsubscript{2} fi} results in executing \texttt{S\textsubscript{1}} if \texttt{C} evaluates to $\true$ and otherwise it executes the statement \texttt{S\textsubscript{2}}. 

\begin{figure}[!t]
\centering
\includegraphics[scale=0.6]{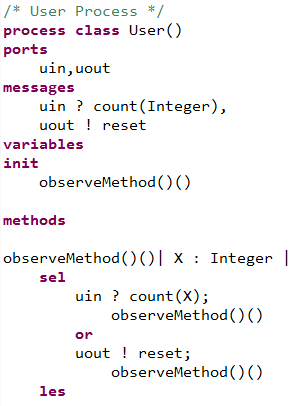}
\caption{A Process Specification for Communicating with Instances of \texttt{Counter}\label{fig:acq_user}}
\end{figure}

\begin{figure}[!bp]
\centering
\includegraphics[scale=0.6]{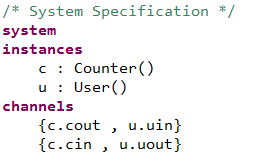}
\caption{A Communication Channel between Instances of \texttt{Counter} and \texttt{User}\label{fig:acq_sys_spec}}
\end{figure}

The process specification in Fig.~\ref{fig:acq_counter_poosl} declares two methods, namely, \texttt{initialize} and \texttt{countMethod}; instances of \texttt{Counter} initially behave as \texttt{initialize}. The method \texttt{initialize} sets \texttt{n} to zero and calls \texttt{countMethod}. A combination of non-deterministic choices and guarded commands are used for case distinction in \texttt{countMethod}. If \texttt{n<3} holds, the process sends a \texttt{count} message that carries the current number. Afterwards, it increments the current number by one and calls \texttt{countMethod} to continue the counting. If \texttt{n==3} holds, the process receives \texttt{reset} and resets the current number to zero. The method \texttt{countMethod} is called to restart the counting.

As indicated earlier, sending and receiving messages in POOSL should be performed in a synchronous way. For instance, in our example, \texttt{cout!count(n)} is only executable, if another process is ready to execute a complementary message, e.g., \texttt{uin?count(X)}, such that \texttt{uin} is connected to \texttt{cout} via a channel. By synchronizing these two messages the value of \texttt{n} is bound to \texttt{X}. Similarly, executing \texttt{cin?reset} require a complementary send message from another process. To this end, we specify a process class called \texttt{User} (see Fig.~\ref{fig:acq_user}) to exchange messages with \texttt{Counter}.

The behavior of \texttt{User} is described in terms a method called  \texttt{observeMethod}. The non-deterministic choice in \texttt{observeMethod} allows instances of the process to receive \texttt{count} and send \texttt{reset} when the complementary messages are available.

\begin{figure}[!tp]
\centering
\includegraphics[scale=0.57]{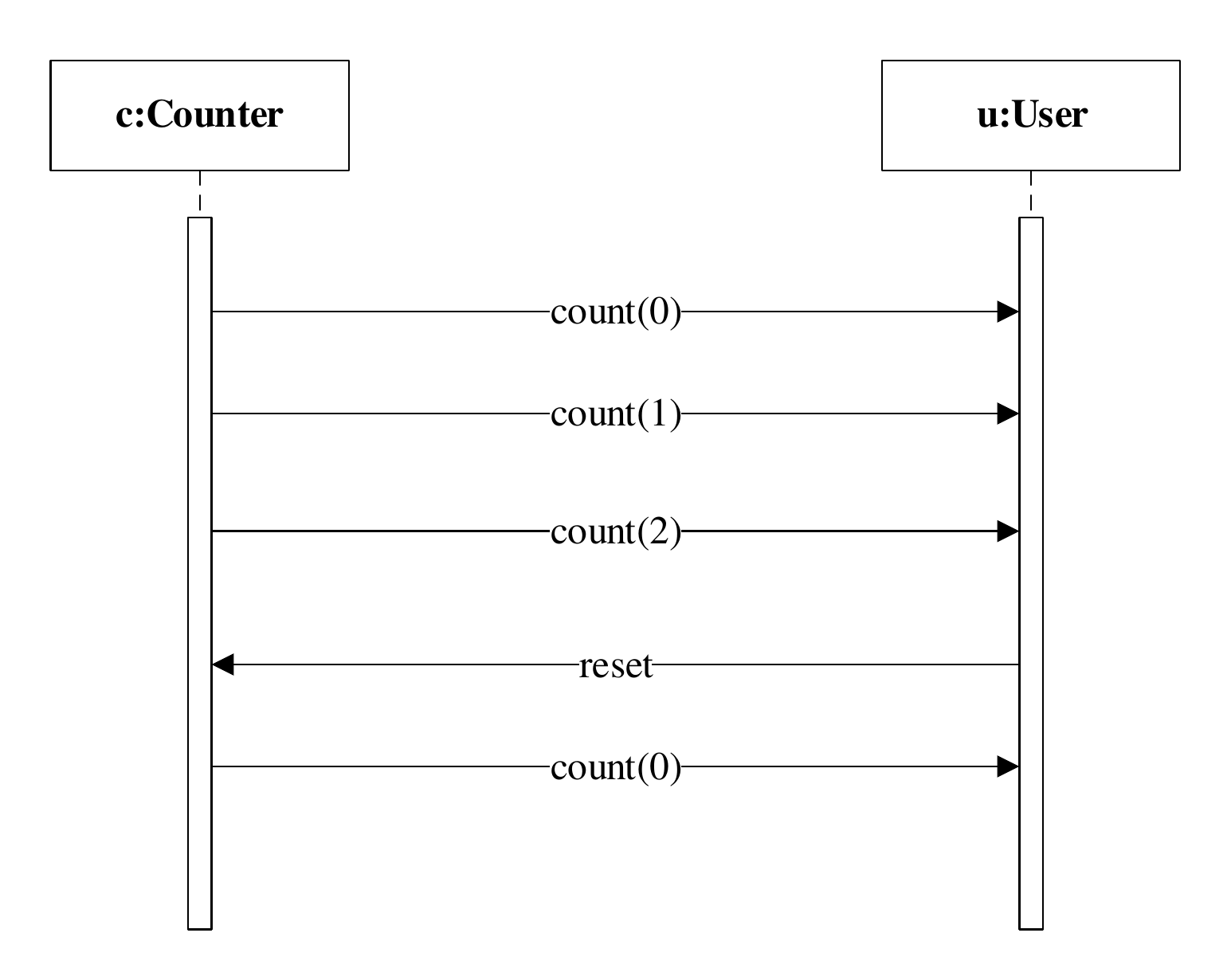}
\caption{Simulating the Interactions between a \texttt{Counter} and \texttt{User}\label{fig:acq_seq}}
\end{figure}

Finally, to obtain an executable model for the modulo $3$ counter, we need concrete instances of \texttt{Counter} and \texttt{User} which are connected via suitable channels. The system specification from Fig.~\ref{fig:acq_sys_spec} instantiates the defined process classes and creates two communication channels between the instances by connecting their ports. Fig.~\ref{fig:acq_seq} illustrates a sequence diagram obtained by simulating the resulting model; the diagram consists of two lifelines (for the instances \texttt{c} and \texttt{u} of the process classes) and depicts the first five steps of their communication.

\subsection{A Transformation from the DSL to POOSL}\label{subsec:acq_poosl_trans}
In this section we describe our transformation scheme for generating POOSL models from DSL models. Similar to the example model from Section~\ref{subsec:acq_poosl_intro}, simulation models generated from the DSL consist of two process classes: 
\begin{itemize}
\item
\texttt{DecisionClass}: receives the inputs declared in the DSL model and makes decisions about the output X-ray based on the rules specified in the DSL model;
\item
\texttt{User}: sends the inputs declared in the DSL model to \texttt{DecisionClass} and observes the produced outputs.
\end{itemize}
In Section~\ref{subsec:acq_poosl_intro}, we specified a simple process class for observing the behavior performed by instances of \texttt{Counter}. A similar structure can be used to specify the process class \texttt{User} to observe the behavior described by a given DSL model. Thus, in this section we focus on the way we obtain \texttt{DecisionClass} from a given DSL model. We use the DSL model from Fig.~\ref{fig:acq_dslexample} to illustrate the way the domain concepts identified in Section~\ref{sec:acq_semantics} are described in POOSL. 

\begin{figure}[!tp]
\centering
\includegraphics[scale=0.6]{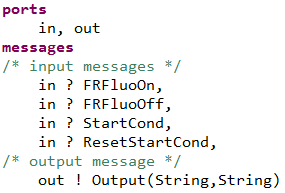}
\caption{Declaring Input and Output Messages in POOSL for the Example DSL Model}\label{fig:acq_poosl_message}
\end{figure}

\paragraph{Action}
In our transformation, port and message declarations of \texttt{DecisionClass} are based on the distinction between the inputs and outputs of \emph{Pedal Handling}. We declare two ports, namely, \texttt{in} (for receiving inputs) and \texttt{out} (for sending outputs); see the port and message declarations of Fig.~\ref{fig:acq_poosl_message} for the DSL model from Fig.~\ref{fig:acq_dslexample}.

Messages that are communicated through the port \texttt{in} correspond to the input actions declared by \texttt{InActions} in a DSL model. For the DSL model from Fig.~\ref{fig:acq_dslexample} we declare \texttt{FRFluoOn,FRFluoOff,StartCond,ResetStartCond} as messages that travel across the port \texttt{in}; see Fig.~\ref{fig:acq_poosl_message}.

As indicated earlier, the output actions of \emph{Pedal Handling} are not explicitly declared in the DSL. The process \texttt{DecisionClass} sends \texttt{Output} on the port \texttt{out} to model the outputs performed by \emph{Pedal Handling}. In comparison with the mCRL2 formalization of Section~\ref{sec:acq_semantics}, we do not use custom data types in POOSL to describe the types of X-ray and the planes; see the declaration of $\xray$ and $\plane$ in Section~\ref{sec:acq_semantics}. Using custom data types in POOSL requires defining new data classes and manipulating their instances. To avoid such complexities in our transformations, we declare \texttt{Output} as an action with two string parameters (see Fig.~\ref{fig:acq_poosl_message}); performing \texttt{Output("Fluo","FR")} indicates a request for X-ray type ${\it Fluo}$ to plane ${\it FR}$. This restricted way of using string parameters is consistent with the use of $\xray$ and $\plane$ as the types of parameters carried by output actions in mCRL2. 

\begin{figure}[b]
    \begin{minipage}{.5\linewidth}
      \centering
        \begin{tabular}{| c | c |}
        \hline
            \textbf{Plane} & \textbf{Integer Value} \\ \hline
            ${\it None}$ & 1 \\ \hline
            ${\it FR}$ & 2 \\ \hline
            ${\it LT}$ & 3 \\ \hline
            ${\it BI}$ & 4 \\ \hline      
        \end{tabular}
    \end{minipage}%
    \begin{minipage}{.5\linewidth}
      \centering
        \begin{tabular}{| c | c |}
        \hline
            \textbf{X-ray Type} & \textbf{Integer Value} \\ \hline
            ${\it Standby}$ & 5 \\ \hline
            ${\it Fluo}$ & 6 \\ \hline
            ${\it SingleShot}$ & 7 \\ \hline
            ${\it Series}$ & 8 \\ \hline
        \end{tabular}
    \end{minipage}
    \caption{Representing Planes and X-ray Types in POOSL}
    \label{fig:acq_map}
\end{figure}

\paragraph{Plane, X-Ray Type} In Section~\ref{sec:acq_semantics} we defined custom data types to describe the planes (i.e, ${\it None}$, ${\it FR}$, ${\it LT}$, ${\it BI}$) and the X-ray types (i.e., ${\it Standby}$, ${\it Fluo}$, ${\it SingleShot}$, ${\it Series}$). As mentioned above, for communicating outputs, we use strings to represent these elements. Since in object-oriented languages performing basic operations on strings, e.g.,  ``equality checking'',  often leads to confusing interpretations such as ``checking equality of values'' and ``checking equality of referenced objects'', we prefer to have a simple way of representing the planes and X-ray types in the internal computations of DSL rules. We represent the elements of these two domain concepts by the sets $\{1,2,3,4\}$ and $\{5,6,7,8\}$, respectively. Fig.~\ref{fig:acq_map} depicts the mapping we apply for representing the planes and X-ray types.

\begin{figure}[t]
\centering
\includegraphics[scale=0.6]{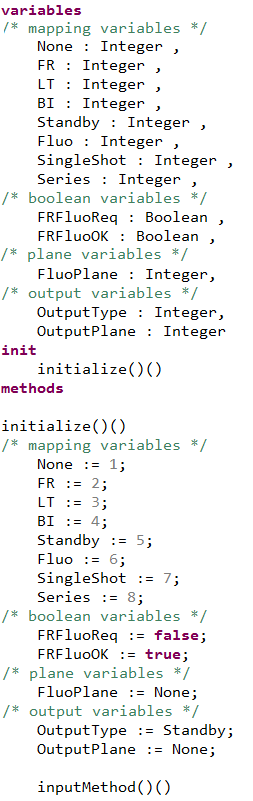}
\caption{Declaring State Variables in POOSL for the Example DSL Model}\label{fig:acq_poosl_variables}
\end{figure}

To represent this mapping in POOSL, we declare an integer variable for each element depicted in the tables of Fig.~\ref{fig:acq_map}; see the variable declarations of Fig.~\ref{fig:acq_poosl_variables}. These variables are initialized based on the defined mapping; the initial values are specified by the method \texttt{initialize} in Fig.~\ref{fig:acq_poosl_variables}. 

To ensure that the mapping is correctly used, the values of these variables are not changed by the methods of the process class after initialization.

\paragraph{State}
As discussed in Section~\ref{sec:acq_semantics}, the notion of state in the DSL is defined based on the set of variables declared in a DSL model
(boolean and plane variables) and 
the two not-explicitly declared output variables (\texttt{OutputType} and \texttt{OutputPlane}).
The declaration of boolean variables in POOSL corresponds to the boolean variables of the DSL model; see the declarations of boolean variables in Fig.~\ref{fig:acq_poosl_variables} for the DSL model from Fig.~\ref{fig:acq_dslexample}. Plane variables and the non-explicitly declared output variables (\texttt{OutputType} and \texttt{OutputPlane}) are denoted by integer variables in POOSL; see the declarations of integer variables in Fig.~\ref{fig:acq_poosl_variables} for the DSL model from Fig.~\ref{fig:acq_dslexample}. 

\begin{figure}[t]
\centering
\includegraphics[scale=0.6]{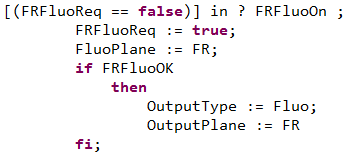}
\caption{Describing a Rule from the Example DSL Model in POOSL}\label{fig:acq_poosl_rule}
\end{figure}

The initial values of the boolean and plane variables are specified in the \texttt{initialize} method and correspond to the initial values described in the DSL model. The output variables are initialized based on the assumption that in the initial state no X-ray is generated from the planes.   


\paragraph{Guard, Do Clause, Rule}
A DSL rule refers to an input action and consists of a guard (specifying when the action is active) and a do clause (specifying the effects of performing the input action). We describe each rule by a guarded command \texttt{[C]S} where \texttt{C} is the guard of the DSL rule and \texttt{S} describes the do clause. Fig.~\ref{fig:acq_poosl_rule} depicts a specification of the first rule from Fig.~\ref{fig:acq_dslexample}.

The do clause of a DSL rule is a sequence of assignments/conditionals. As discussed in Section~\ref{subsec:acq_poosl_intro}, \texttt{x:=E} and \texttt{if C then S\textsubscript{1} else S\textsubscript{2} fi} are among the statement types that can appear in the body of a POOSL method. Thus, we use these constructs to specify do clauses in POOSL. The do clause of the first rule from Fig.~\ref{fig:acq_dslexample} is translated to four assignments and one conditional in POOSL; see the statements guarded by \texttt{[(FRFluoReq==false)]} in Fig.~\ref{fig:acq_poosl_rule}.

\begin{figure}[!t]
\centering
\includegraphics[scale=0.56]{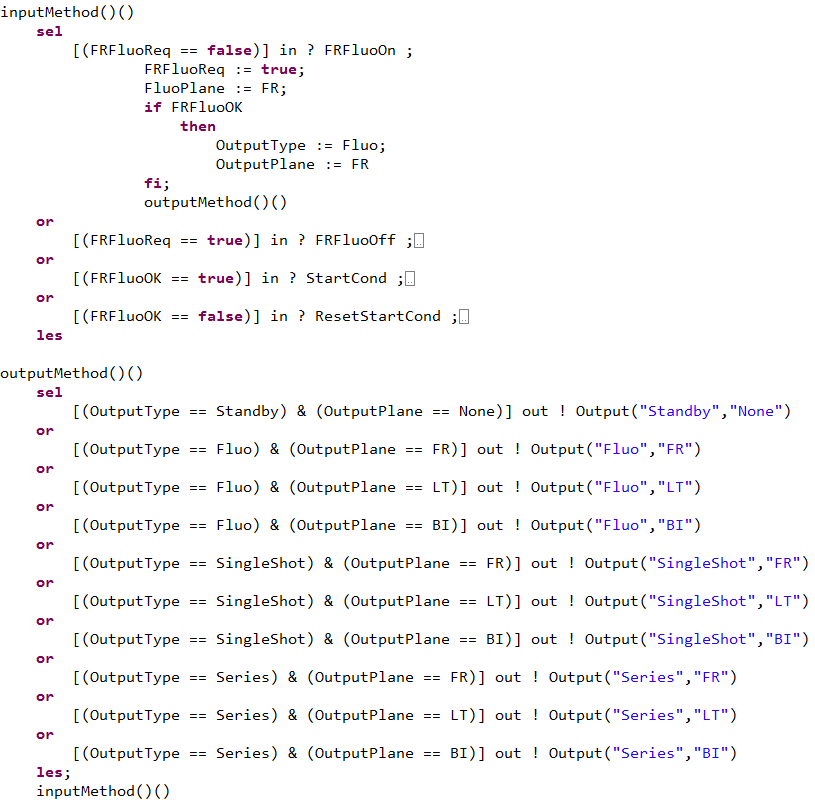}
\caption{POOSL Methods for Receiving Inputs and Producing Outputs}\label{fig:acq_poosl_in_out}
\end{figure}

We repeat this transformation scheme for all the rules of a DSL model and specify a method called \texttt{inputMethod} that describes the way input actions are received by \emph{Pedal Handling}. Fig.~\ref{fig:acq_poosl_in_out} depicts this transformation for the DSL model from Fig.~\ref{fig:acq_dslexample}. The body of \texttt{inputMethod} consists of a non-deterministic choice between the guarded commands that describe the DSL rules. Note that in Fig.~\ref{fig:acq_poosl_in_out} we have folded the descriptions of the last three rules. 

Observe that in Fig.~\ref{fig:acq_poosl_in_out} we have extended the description of Fig.~\ref{fig:acq_poosl_rule} for the first rule of the DSL model. A sequential composition of the do clause and \texttt{outputMethod} is used to show that after computing the effects of the action, \emph{Pedal Handling} is ready to produce an output. Each folded specification block also consists of a sequential composition of the corresponding do clause and \texttt{outputMethod}.

The method \texttt{outputMethod} describes the behavior of \emph{Pedal Handling} when the component is ready to perform an output action. A combination of non-deterministic choices and guarded commands are applied to transform the integer values of \texttt{OutputType} and \texttt{OutputPlane} into string parameters for the action \texttt{Output} (see the mapping from Fig.~\ref{fig:acq_map}). After performing an output action the method \texttt{inputMethod} is called and the model can receive the next input. 

Similar to the process $P_{\din}$ and $P_{\dout}$ in the mCRL2 formalization, we use the methods \texttt{inputMethod} and \texttt{outputMethod} to enforce alternative execution of the input and output actions.

\section[Validating the Industrial Implementation by Model-based Testing]{Validating the Industrial Implementation by\\ Model-based Testing}
\label{sec:acq_mbt}

In industry, the implementation is considered to be the most valuable artifact that is produced for a component. When behavioral models of a component are also available, it is relevant to check whether the implementation complies to the modeled behavior. Validating the compliance of implementations to DSL models adds a level of redundancy that can reveal discrepancies between the developed DSL models and implementations. 

We use model-based testing to validate the correctness of an implementation with respect to a DSL model. Model-based testing uses a model that describes the behavior of a system under test and enables both automatic generation and execution of test cases on the implementation. 

The semantics of the pedal handling DSL is described in terms of labeled transition systems and hence we use the theory of input-output conformance (\textit{ioco}) testing for labeled transition systems \cite{T08} to automatically derive test cases from the behavior described in the DSL and execute them on the implementation. 

In the \textit{ioco} theory, correctness of implementations with respect to specifications is expressed in terms of the binary relation of \textit{ioco}. The \textit{ioco} theory provides an algorithm that derives a set of test cases from a given specification such that executing this set of test cases on an implementation determines whether the specification and implementation are related by the conformance relation. 

The model-based testing tool of Axini \cite{Axini} is based on the \textit{ioco} theory. In this tool, specifications are described using the Axini Modeling Language (AML). In order to facilitate model-based testing on implementations of the \emph{Pedal Handling} component, we aim to develop an automated transformation from the DSL to AML (Fig.~\ref{fig:acq_artifacts}).

In this section first we introduce the AML language (Section~\ref{subsec:acq_aml}) and describe our transformation from the DSL to AML (Section~\ref{subsec:acq_dsl_aml}). Then we discuss about model-based testing of the industrial implementation using the generated models and interpreting the obtained results (Section~\ref{subsec:acq_interpret}). Finally, we discuss some practical remarks (Section~\ref{subsec:acq_practical}).

\subsection{An Overview of AML}\label{subsec:acq_aml}
AML is a modeling language that allows the modeler to formally specify the behavior of a system that communicates with its environment through sending and receiving actions. Symbolic labeled transition systems is the formalism used by Axini for giving semantics to the language constructs. In what follows, we give a brief overview of the syntax of AML\footnote{AML is a modeling language designed by Axini and is only available as part of their commercial model-based testing toolset. For confidentiality reasons, we use a modified version of the language in this paper.}; our focus is on the language constructs that we need throughout this paper for constructing test models from specifications in the \emph{Pedal Handling} DSL. We use the modulo $3$ counter from Section~\ref{sec:acq_semantics} as a running example to discuss the constructs of AML.

In AML behavior is described in terms of processes. An AML process starts by declaring the inputs actions (stimuli) that it can accept and the observable output actions (responses) that it can produce. For instance, the following AML statements declare a stimulus (called \texttt{reset})  and a response (called \texttt{count}) for the modulo $3$ counter: 
\begin{Verbatim}[commandchars=\\\{\}]
  stimulus \sq\hspace{0pt}reset\sq
  response \sq\hspace{0pt}count\sq , \sq\hspace{0pt}curr\sq:integer
\end{Verbatim}

For each action, the direction of communication should be denoted by indicating the keyword \texttt{stimulus} (input) or \texttt{response} (output). Actions can carry data parameters. For instance, in the example above \texttt{count} carries an integer parameter called \texttt{curr} to indicate the current number. 

Action declarations in AML processes is followed by declaring state variables. State variables of a process are data items that determine the current state of the process and can influence the course of actions. AML supports various data types for variables, e.g., booleans, integer numbers, strings. A variable declaration in AML consists of a name, a data type, and the initial value of the variable. For the modulo $3$ counter, we declare an integer variable called \texttt{n} to keep track of the current number. The following AML statement declares this variable and initializes it to zero:
\begin{Verbatim}[commandchars=\\\{\}]
  var \sq\hspace{0pt}n\sq, :integer, 0
\end{Verbatim}
  
Assigning values to state variables and receiving/sending actions are among the basic operations that can be performed by a process. The following list provides examples of these basic operations for the modulo $3$ counter:
\begin{itemize}
\item
\texttt{update \sq\hspace{0pt}n=0\sq}: resetting the state variable \texttt{n} to zero;
\item
\texttt{receive \sq \hspace{0pt}reset\sq}: receiving the \texttt{reset} action from the environment;
\item
\texttt{send  \sq\hspace{0pt}count\sq, constraint: \sq\hspace{0pt}curr==n\sq}: sending the \texttt{count} action to report the current number to the environment.
\end{itemize}
The \texttt{send} example shows that for actions carrying data parameters we can specify a  \texttt{constraint}, which is a boolean expression, to describe the desired values that are communicated with the environment. 

AML also provides operators that combine the basic operations and allow us to specify more complex behaviors. Suppose \texttt{S\textsubscript{1},S\textsubscript{2}} are AML statements. The non-deterministic choice between 
\texttt{S\textsubscript{1}} and \texttt{S\textsubscript{2}} is denoted as follows:
\begin{Verbatim}[commandchars=\\\{\}]
  choice \{ 
  \quad\quad o \{ S\textsubscript{1} \} 
  \quad\quad o \{ S\textsubscript{2} \}
   \}
\end{Verbatim}

A specification that consist of \texttt{S\textsubscript{1}} and \texttt{S\textsubscript{2}} with a line break in between describes the sequential composition of \texttt{S\textsubscript{1}} and \texttt{S\textsubscript{2}}. 

The values of state variables in a process can also influence the course of actions in the process. Suppose that \texttt{S} is an AML statement and \texttt{C} is a boolean expression defined in terms of the state variables of a process. The guarded statement \texttt{guard \sq\hspace{0pt}C\sq \hspace{0.5pt} S} only allows the execution of \texttt{S} if the condition \texttt{C} hold. If \texttt{C} evaluates to $\false$, the guarded statement \texttt{guard \sq\hspace{0pt}C\sq\hspace{0.5pt} S} will lead to a deadlock. 

In Fig.~\ref{fig:acq_counter_aml}, we apply the constructs discussed above to specify the modulo $3$ counter in AML. Note that the behavioral description is enclosed in a block labeled by \texttt{ready}. This block consists of a non-deterministic choice between two disjoint cases, namely, \texttt{n<3} and \texttt{n==3}. In the first case, \texttt{count} is performed and the current number is incremented by $1$, whereas in the second case \texttt{reset} is performed and \texttt{n} becomes zero. In both cases, the \texttt{goto} statement allows us to repeat the behavior defined in the \texttt{ready} block. 

\begin{figure}[t]
\begin{Verbatim}[commandchars=\\\{\}]
   process(\sq\hspace{0pt}Counter\sq) \{
		
       stimulus \sq\hspace{0pt}reset\sq
       response \sq\hspace{0pt}count\sq,\sq\hspace{0pt}curr\sq => :integer
		
       var \sq\hspace{0pt}n\sq, :integer, 0
		
       label \sq\hspace{0pt}ready\sq
           choice \{
            o \{
              guard \sq\hspace{0pt}n<3\sq
                 send \sq\hspace{0pt}count\sq, constraint: \sq\hspace{0pt}curr==n\sq
                 update \sq\hspace{0pt}n=n+1\sq
                 goto \sq\hspace{0pt}ready\sq
              \}
            o \{
              guard \sq\hspace{0pt}n==3\sq
                 receive \sq\hspace{0pt}reset\sq
                 update \sq\hspace{0pt}n=0\sq
                 goto \sq\hspace{0pt}ready\sq
              \}
       \}	
   \}
\end{Verbatim}
\caption{A Process Specification of the Modulo $3$ Counter in AML\label{fig:acq_counter_aml}}
\end{figure}

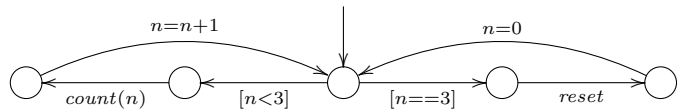
\begin{figure}[!t]
\centerline{
\xymatrix@C=4pc{
&&\ar[d]&&\\
*++[o][F]{}\ar@/^1.3pc/[rr]^{n=n+1}&*++[o][F]{}\ar[l]^{{\it count}(n)}&*++[o][F]{}\ar[l]^{[n<3]} \ar[r]_{[n==3]} &*++[o][F]{} \ar[r]_{{\it reset}}&*++[o][F]{}\ar@/_1.3pc/[ll]_{n=0}
}
}
\caption{A Symbolic Labeled Transition System Describing the Behavior of \texttt{Counter}}
\label{fig:acq_aml_lts}
\end{figure}

For a given AML model, the toolset of Axini can generate a symbolic labeled transition system that corresponds to the behavior described in the model.  Fig.~\ref{fig:acq_aml_lts} depicts the behavior of \texttt{Counter} in terms a symbolic labeled transition system; to distinguish between the guards and assignments, we put the guards between square brackets.

\subsection{A Transformation from the DSL to AML}\label{subsec:acq_dsl_aml}
In this section we consider the domain concepts identified in Section~\ref{sec:acq_semantics} and discuss our transformation scheme for generating AML models from models in the \emph{Pedal Handling} DSL. We use the DSL model from Fig.~\ref{fig:acq_dslexample} to describe our transformation. 
\paragraph{Action}
For a given DSL model, we declare two types of actions in AML:
\begin{itemize}
\item
stimuli (input actions), correspond to the actions declared by \texttt{InActions} in the DSL model;
\item
response (output action), an action  carrying two string parameters that indicate an X-ray request for a specific plane.
\end{itemize}
For instance, for the DSL model from Fig.~\ref{fig:acq_dslexample}, this leads to the following action declarations:
\begin{Verbatim}[commandchars=\\\{\}]
stimulus \sq\hspace{0pt}FRFluoOn\sq,\sq\hspace{0pt}FRFluoOff\sq,\sq\hspace{0pt}StartCond\sq,\sq\hspace{0pt}ResetStartCond\sq
response \sq\hspace{0pt}Output\sq,\sq\hspace{0pt}xraytype\sq=>:string,\sq\hspace{0pt}xraychannel\sq =>:string
\end{Verbatim}
When sending the action \texttt{Output}, we can use the parameter names \texttt{xraytype} and \texttt{xraychannel} to specify constraints on the data values that are sent through this action.  

\paragraph{Plane, X-Ray Type}
In Section~\ref{sec:acq_semantics} we used custom data types to describe the planes and X-ray types. Although the non-disclosed parts of AML allow us to define custom data types, we have decided to use a limited set of constructs from AML and to represent these two domain concepts with strings. 

Strings are a predefined data type in AML. Since AML is not an object-oriented language, accessing strings by reference is not a relevant issue in the language. All operations on strings directly manipulate the values. Thus, we use:
\begin{itemize}
\item
\texttt{"None"}, \texttt{"FR"}, \texttt{"LT"}, \texttt{"BI"} to represent the planes;
\item
\texttt{"Standby"}, \texttt{"Fluo"}, \texttt{"SingleShot"}, \texttt{"Series"} to represent the X-ray types.  
\end{itemize}

\paragraph{State}
For a given DSL model, the notion of state is defined based on the declared boolean variables, plane variables, and the two non-explicitly declared output variables (\texttt{OutputType} and \texttt{OutputPlane}). In AML this notion of state is described in terms of booleans (representing the boolean variables of the DSL model) and string variables (representing the plain and output variables of the DSL model).

For the DSL model from Fig.~\ref{fig:acq_dslexample} with two boolean variables (i.e., \texttt{FRFluoReq} and \texttt{FRFluoOK}) and one plane variable (i.e., \texttt{FluoPlane}), we use the following declarations in AML (statements preceded by \texttt{\#} are comments):
\begin{Verbatim}[commandchars=\\\{\}]
#boolean variables
var \sq\hspace{0pt}FRFluoReq\sq, :boolean, false
var \sq\hspace{0pt}FRFluoOK\sq, :boolean, true

#plane variables
var \sq\hspace{0pt}FluoPlane\sq, :string, "None"

#output variables
var \sq\hspace{0pt}OutputChannel\sq, :string, "None"
var \sq\hspace{0pt}OutputType\sq, :string, "Standby"
\end{Verbatim}
The declared variables are also initialized; the initialization of the boolean and plane variables are based on the initial values specified by \texttt{init} in the DSL model (see Fig.~\ref{fig:acq_dslexample}). The initial values of \texttt{OutputChannel} and \texttt{OutputType} are based on the domain assumption that initially no X-ray is generated from the planes.

\paragraph{Guard, Do Clause, Rule}
A DSL model specifies one rule for each input action to indicate when the action is executable (i.e., guard of the rule) and the way performing the action influences the current state (i.e., do clause of the rule). We transform each rule to a guarded statement \texttt{guard \sq\hspace{0pt}C\sq\hspace{0.5pt} S} in AML such that \texttt{C} is the guard of the rule and \texttt{S} describes the do clause.

Since do clauses consist of assignments and conditionals, we have to provide transformation schemes for these two constructs. Assignments are transformed to \texttt{update} statements in AML. A conditional statement \texttt{if C then S fi} from the DSL is transformed to a non-deterministic choice with two options as follows:
\begin{Verbatim}[commandchars=\\\{\}]
choice \{
   o \{ guard \sq\hspace{0pt}C\sq  S\}
   o \{ guard \sq\hspace{0pt}!C\sq  \}
\}   
\end{Verbatim}
The first option allows the execution of \texttt{S}, if \texttt{C} evaluates to $\true$. The second option is a guarded statement with an empty body (i.e., skip) and gets executed  if \texttt{!C}  holds. 

The following AML statements describe the first rule of the DSL example from Fig.~\ref{fig:acq_dslexample}:
\begin{Verbatim}[commandchars=\\\{\}]
guard \sq\hspace{0pt}(FRFluoReq == false)\sq
   receive \sq\hspace{0pt}FRFluoOn\sq                 
   update \sq\hspace{0pt}FRFluoReq=true\sq            # do clause assignment
   update \sq\hspace{0pt}FluoPlane="FR"\sq            # do clause assignment
   choice \{                           # do clause conditional 
     o \{
       guard \sq\hspace{0pt}FRFluoOK\sq               # then part of conditional
         update \sq\hspace{0pt}OutputType="Fluo"\sq   # do clause assignment
         update \sq\hspace{0pt}OutputChannel="FR"\sq  # do clause assignment 
      \}
     o \{
       guard \sq\hspace{0pt}!(FRFluoOK)\sq
      \}
   \}
\end{Verbatim}
The do clause of the first rule from Fig.~\ref{fig:acq_dslexample} consists of four assignments and one conditional statement. As illustrated above, the assignments are transformed to updates and the conditional statement is described in terms of a choice in AML. 

For a given DSL model, we apply this transformation scheme to each rule of the model and use the obtained AML fragments to construct a process specification that describes the behavior specified in the DSL model. Fig.~\ref{fig:acq_aml} depicts the process specification that we obtain by applying our transformation to the DSL model from Fig.~\ref{fig:acq_dslexample}; we have folded the specification blocks of the last three rules.

\begin{figure}
\begin{Verbatim}[commandchars=\\\{\}]
 process(\sq\hspace{0pt}DecisionProcess\sq) \{		
   stimulus \sq\hspace{0pt}FRFluoOn\sq,\sq\hspace{0pt}FRFluoOff\sq,\sq\hspace{0pt}StartCond\sq,\sq\hspace{0pt}ResetStartCond\sq
   response \sq\hspace{0pt}Output\sq,\sq\hspace{0pt}xraytype\sq=>:string,\sq\hspace{0pt}xraychannel\sq =>:string
     
   #boolean variables
   var \sq\hspace{0pt}FRFluoReq\sq, :boolean, false
   var \sq\hspace{0pt}FRFluoOK\sq, :boolean, true

   #plane variables
   var \sq\hspace{0pt}FluoPlane\sq, :string, "None"

   #output variables
   var \sq\hspace{0pt}OutputChannel\sq, :string, "None"
   var \sq\hspace{0pt}OutputType\sq, :string, "Standby"
     
   label \sq\hspace{0pt}ready\sq  
   choice \{
    o \{ #Rule 1
      guard \sq\hspace{0pt}(FRFluoReq == false)\sq
        receive \sq\hspace{0pt}FRFluoOn\sq                 
        update \sq\hspace{0pt}FRFluoReq=true\sq         # do clause assignment
        update \sq\hspace{0pt}FluoPlane="FR"\sq         # do clause assignment
        choice \{                        # do clause conditional 
          o \{
           guard \sq\hspace{0pt}FRFluoOK\sq             # then part of conditional
           update \sq\hspace{0pt}OutputType="Fluo"\sq   # do clause assignment
           update \sq\hspace{0pt}OutputChannel="FR"\sq  # do clause assignment 
          \}
          o \{
           guard \sq\hspace{0pt}!(FRFluoOK)\sq
          \}
          
          # performing output action
          send \sq\hspace{0pt}Output\sq, constraint: \sq\hspace{0pt}xraytype==OutputType && 
                                      xraychannel==OutputChannel\sq
          #return to \sq\hspace{0pt}ready\sq                             
          goto \sq\hspace{0pt}ready\sq
     \}
         
    o \{ #Rule 2 \}  
    o \{ #Rule 3 \}
    o \{ #Rule 4 \}          
   \} 
 \}
\end{Verbatim}
\caption{An AML Process Specification for the Example DSL Model\label{fig:acq_aml}}
\end{figure}

In Fig.~\ref{fig:acq_aml} the behavioral specification is encloses in a block labeled by \texttt{\sq\hspace{0pt}ready\sq}. This block consists of a non-deterministic choice between the rules of the DSL model. The guarded statement inside each option describes a rule and only allows the rule to be chosen if it is active (i.e. the guard evaluates to $\true$). 

The first option in the non-deterministic choice describes the first rule from Fig.~\ref{fig:acq_dslexample}. Observe that we have extended the specification of the first rule with two sequential compositions. This indicates that after evaluating the do clause an output action is performed and then the behavior of the \texttt{\sq\hspace{0pt}ready\sq} block is repeated. The folded blocks also have these sequential compositions as their last steps. This enforces the alternating execution of the input and output actions as specified by the processes $P_{\din}$ and $P_{\dout}$ in Section~\ref{sec:acq_semantics}.

\subsection{Interpreting the Results of Model-based Testing}\label{subsec:acq_interpret}
Validating the compliance of an implementation to a DSL model by model-based testing requires extra care to interpret the results correctly. A failed test case shows a discrepancy between the behavior described by the test model and the behavior implemented in the implementation. This can have two different reasons:
\begin{itemize}
\item
a mistake in the test models generated from the DSL;
\item
a failure in the implementation.
\end{itemize}
The transformation from the DSL to test models should preserve the semantics of the DSL. A mistake in realizing the semantics may result in failed test cases. In Section~\ref{sec:acq_comparison} we introduce an approach to gain more confidence in the correctness of the generated models in Fig.~\ref{fig:acq_artifacts}. When there is sufficient confidence in the correctness of test models, a failed test case indicates a failure of the implementation.

It should be noted that if implementations are not generated from the DSL, failed test cases could also indicate a mistake in DSL models; the intended behavior is not correctly described in the DSL but the implementation has correctly realized the behavior.

\subsection{Practical Remarks}\label{subsec:acq_practical}
\begin{figure}[!t]
\centering
\includegraphics[scale=0.53]{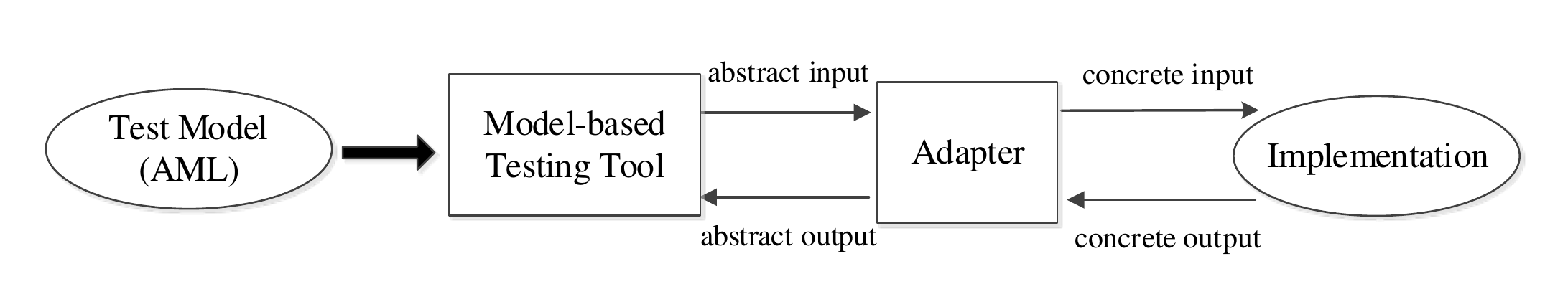}
\caption{Model-based Testing}\label{fig:acq_mbt}
\end{figure}

Fig.~\ref{fig:acq_mbt} depicts the environment we have developed for model-based testing. The tool of Axini uses an on-the-fly algorithm for test case generation and execution. This means that in each step of a test case the tool can do one of the following:
\begin{itemize}
\item
send an input action allowed by the model to the implementation;
\item
wait for an output from the implementation and check that the produced output is allowed by the model.
\end{itemize}

The adapter in Fig.~\ref{fig:acq_mbt} translates the abstract actions from the model to concrete commands for the implementation and maps the concrete outputs produced by the implementation to abstract output actions for the model.

\paragraph{Results}
In our model-based testing experiments, we encountered failed test cases that indicated a failure in a not-yet-released implementation which is not generated from the DSL. For example, one of the requirements of \textit{Pedal Handling} indicates that high-dose X-ray requests have priority over low-dose X-ray requests. A design mistake in realizing this requirement led to a failed test case. The failed test case was part of an unlikely scenario where three different pedals must be pressed at the same time.

Model-based testing revealed that certain modeling choices taken in our DSL models are implemented differently in the implementation. Unlike stopping \emph{Fluo} (modeled by \textit{FRFluoOff} in the example of Fig.~\ref{fig:acq_dslexample}), stopping \emph{Series} requires performing two actions in a specific order. The output specified for the first step of stopping \emph{Series} was different from the output produced by the implementation.  

\section[Checking the Correctness of Generated Models
through Redundancy]{Checking the Correctness of Generated \\ Models through Redundancy}
\label{sec:acq_comparison}
Transformations from a DSL to analysis models allow the user to apply various formal techniques and reason about DSL models. Analysis models can also be used to assess the correctness of other artifacts available for a component (see Section~\ref{sec:acq_mbt}). However, the results obtained from analysis models are only valuable if the corresponding transformations correctly realize the semantics of the language.  

Developing a transformation from a DSL to a modeling language requires a deep understanding of the semantics of the DSL and the target language. Moreover, the transformation should not deviate from the semantics of the DSL. Introducing redundancy is a very effective way to reduce the rate of mistakes in such error-prone tasks \cite{BG13}.

\begin{figure}[!t]
\centering
\includegraphics[scale=0.55]{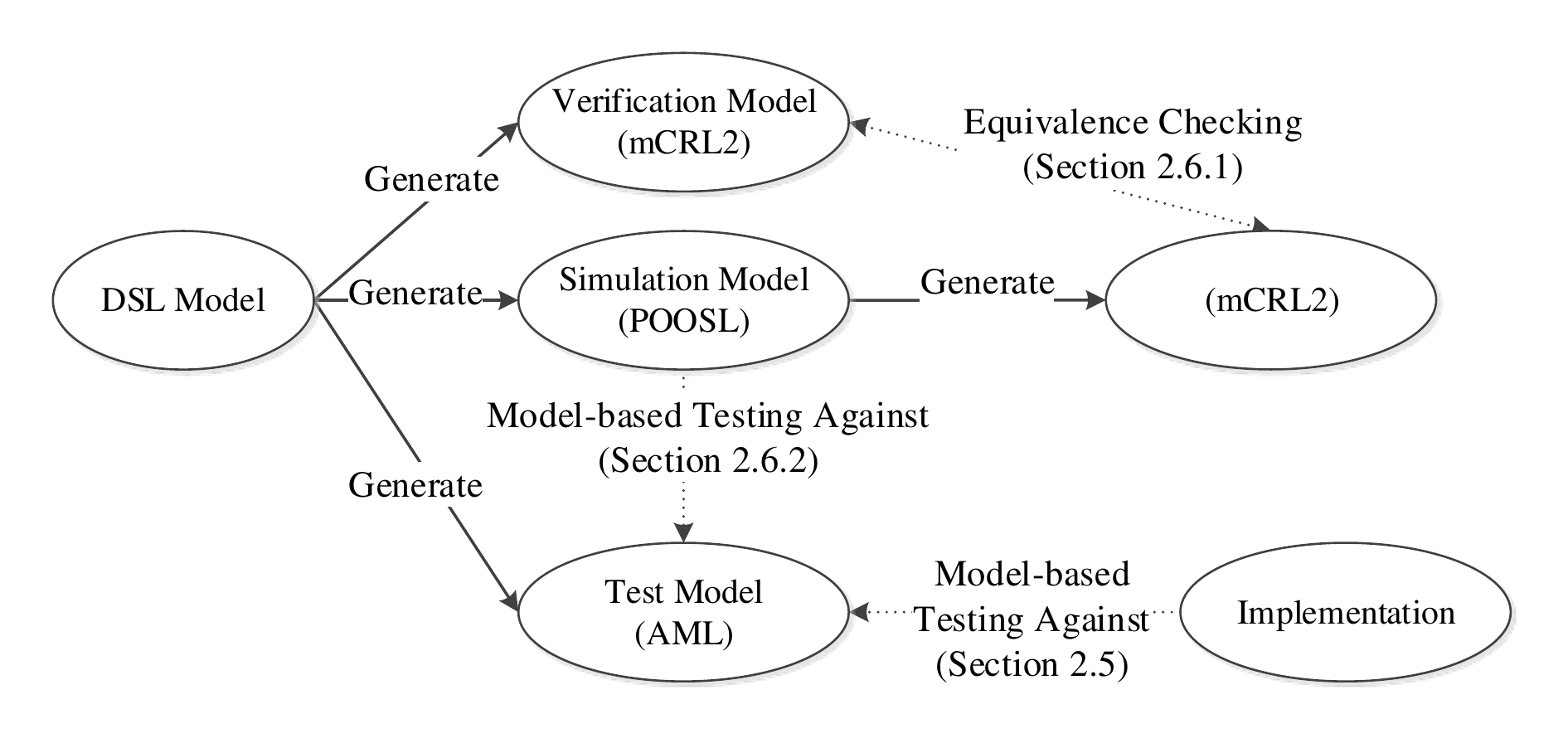}
\caption{Validation of Artifacts\label{fig:acq_redundancy}}
\end{figure}

We introduce redundancy to validate the correctness of the transformations depicted in Fig.~\ref{fig:acq_artifacts} in two ways: equivalence checking (Section~\ref{subsec:acq_red_eqn}) and model-based testing (Section~\ref{subsec:acq_red_mbt}).

%

\subsection{Checking the Behavioral Equivalence between Artifacts}\label{subsec:acq_red_eqn}
The verification and simulation models in Fig.~\ref{fig:acq_redundancy} have an underlying labeled transition system. To get confidence in the correctness of simulation models with respect to the formalized semantics (transformation to verification models), we can investigate whether the labeled transition systems described by the simulation and verification models are related by an equivalence relation, e.g., strong bisimulation. This may require to develop transformations from analysis models to a formalism that enables state space generation and comparison. 

We have used the mCRL2 toolset for state space generation and comparison. To enable state space generation for simulation models, we have used a previously implemented transformation\footnote{The transformation from POOSL to mCRL2 is based on the formal semantics of POOSL represented in \cite{B02} and only supports the syntactic constructs discussed in Section~\ref{sec:acq_poosl}. The transformation is implemented by Arjan J. Mooij.} from POOSL to mCRL2; see Section~\ref{sec:acq_transformations} for some observations regarding this transformation.  

When comparing behaviors, the internal steps performed by them are not relevant; we focus on the observable behaviors. Hence, we check whether the state spaces are equivalent modulo branching bisimulation. Fig.~\ref{fig:acq_redundancy} depicts equivalence checking between mCRL2 and POOSL models and the automated transformation from POOSL to mCRL2 that enables this check. 

\paragraph{Results}
For actual DSL models, the behaviors described in mCRL2 and POOSL were equivalent modulo branching bisimulation. Models generated from three transformations (DSL to mCRL2, DSL to POOSL, and POOSL to mCRL2) are involved in equivalence checking between simulation and verification models. Each transformation is implemented by a different person. This reduces the probability of identical mistakes in the transformations and makes the redundancy introduced by equivalence checking more valuable. 

\subsection[Model-based Testing of Executable Models]{Model-based Testing of Executable Models}
\label{subsec:acq_red_mbt}

In Section~\ref{sec:acq_mbt} we used model-based testing to validate the correctness of an implementation. Executable analysis models can also be treated as black-boxes that interact via their interfaces; we can supply inputs to an executable model and observe its output. Thus, executable models can be tested against the test model generated from a DSL specification using model-based testing. Failed test cases reveal mistakes in the transformations. 

In our case study, we have applied model-based testing to POOSL models; see Fig.~\ref{fig:acq_redundancy}. To this end, we have developed the environment of Fig.~\ref{fig:acq_mbt} with a dedicated adapter for POOSL models.

\paragraph{Results}
We did not encounter any failed test cases in our model-based testing experiments against POOSL models. Similar to Section~\ref{subsec:acq_red_eqn}, the transformations to simulation models (POOSL) and test models (AML) are implemented by different people. This gives more confidence that the semantics of the DSL is realized correctly in these models.

\section{Experiences with Model Transformations}\label{sec:acq_transformations}
The approach of Fig.~\ref{fig:acq_redundancy} deploys model transformations to enable the use of multiple formal techniques and to introduce redundant mechanisms for assessing the correctness of different artifacts with respect to the formalized semantics. This approach relies on two types of transformation: transformations from the DSL to a general-purpose modeling language (from the DSL to mCRL2, POOSL, and AML), and transformations between general-purpose modeling languages (from POOSL to mCRL2). In this section we report on our experiences with these two types of transformations. 

General-purpose modeling languages originate from different disciplines and are applicable to a wide range of problems. For instance, POOSL is designed to be applicable for simulation and performance analysis, whereas mCRL2 is focused on formal verification. In our experience, making transformations between two general-purpose formalisms is not a trivial task. There are usually language constructs from the source language that are difficult or even impossible to translate to the target language. 
 
For example, the data layer of POOSL is object-oriented. Each data class describes its variables and methods.  Moreover, instances of certain data structures (e.g., strings, lists) are accessed using pointers. On the other hand, mCRL2 is not object-oriented and does not support pointers. Hence, transforming data classes or pointers from POOSL would require complex mechanisms in mCRL2 models which in turn would increase the rate of mistakes in the transformation. 

Similarities between the source and target languages may suggest a direct mapping between certain constructs. However, similar constructs in two formalisms may have subtly different semantics. For instance, one would expect the conditional statement \texttt{if c then p else q fi} of POOSL to be trace equivalent to $c\rightarrow p\diamond q$ in mCRL2. Based on the formal semantics of POOSL, \texttt{if c then p else q fi} first performs an internal step ($\tau$ transition) to evaluate the condition. Then it behaves as \texttt{p} if \texttt{c} holds and otherwise it behaves as \texttt{q}. However, in mCRL2 no internal action is performed for evaluating $c$ and hence the two conditional statements are not trace equivalent. 

As another example, we consider  \texttt{[c]p} in POOSL and $c\rightarrow p$ in mCRL2. In both cases, the condition $c$ is evaluated without performing an internal step. For both \texttt{[c]p} and $c\rightarrow p$, if the condition evaluates to $\true$, $p$ is executed and otherwise a deadlock occurs. Thus, \texttt{[c]p} and $c\rightarrow p$ are trace equivalent. These examples of similar constructs from POOSL and mCRL2 reiterate the importance of formal semantics in transformations between languages. 

In the literature, some authors have reported similar experiences about transformations between general-purpose modeling languages. A partial transformation from the hybrid modeling formalism Chi 2.0 \cite{BHRRS08} to mCRL2 is proposed in \cite{S12}. The intention is to enable formal verification on models in Chi 2.0. Semantic differences between Chi 2.0 and mCRL2 makes the transformation complex and hard to maintain. The generated models are also complex and sometimes difficult to analyze by tools. In \cite{HKOVY08} the authors use a model checker and a theorem prover for verification of UML diagrams. Representing the semantics of UML diagrams in these tools and the consistency of these representations are among the main challenges in this work.

The \textit{Pedal Handling} DSL is defined for a narrow domain and its semantics is less elaborate compared with POOSL, mCRL2, and AML. Thus, it requires less effort to construct the transformations from the DSL to different analysis models.

\section{Related Work}\label{sec:acq_related}
In \cite{ABE11} the authors prototype the semantics of a DSL called SLCO. The core semantics is described in terms of a transformation to an intermediate language called CS. Afterwards, CS models are transformed to labeled transition systems and are inspected manually or analyzed by existing tools. During the development of SLCO, the authors have implemented a number of transformations, including transformations from SLCO to SLCO models with equivalent behaviors.
The authors suggest that the prototype semantics can be used to compare the underlying labeled transition systems of the source and target models.

The B method \cite{AAH05} has been used in \cite{BFLM05} to develop process schedulers based on specifications in a DSL. The information given by a DSL model is taken into account at several refinement steps in B machines. The authors also introduce a decidable logic for expressing proof obligations of the refinement steps. This allows them to automatically prove the refinements.

The mentioned studies focus on validating refinement steps. In contrast, we offer various formal techniques to assess the semantic correctness of different types of artifacts.

In \cite{SBSM14} a combination of model-based techniques is used to develop a software bus in a two-phase process. In the first phase, an mCRL2 model of the component is created and validated through simulation. After developing the component, the mCRL2 model is used for model-based testing of the implementation. In the second phase, different properties are verified against the mCRL2 model. The model is improved based on the results and is used for model-based testing against a second implementation. In comparison, our approach is centered around domain-specific models and artifacts generated from them.

In \cite{HMW12} the POOSL language is used to create a requirements model and a preliminary design model for \textit{Pedal Handling}. These models are translated to CSP \cite{R10} and compared using the refinement checker FDR2 \cite{FDR10}. In \cite{HMW12} it is assumed that \textit{Pedal Handling} consists of an input buffer with certain scheduling rules and a component that processes the inputs from the buffer to make decisions about X-ray requests to the tubes. Hence, the scope of the models is wider compared to the work presented in this paper. As a result, refinement checking is not always feasible. In contrast, specifications in the \textit{Pedal Handling} DSL focus on the decision making aspect of the component. This reduces the complexity of the models and makes it feasible to compare behaviors by equivalence checking.

In Section~\ref{sec:acq_mbt}, we have used model-based testing to validate an implementation of a DSL model. As an alternative, one can use a model learning approach. Model learning techniques extract an automaton model for an implementation by systematically performing tests on it and observing its behavior \cite{A87}. The extracted model can be compared with the labeled transition system described by a DSL model. 

\section{Conclusions}\label{sec:acq_conclusion}
A DSL allows us to use models that are naturally aligned with the way domain experts reason about a software component. Existing tools enable language designers to define DSLs and to construct transformations to implementation code and analysis models. However, the semantic correctness of the generated artifacts is usually overlooked.

To resolve conflicting interpretations of a DSL, we use a formal semantics of the language. We also propose to have additional mechanisms to validate the correctness of the generated artifacts with respect to the semantics of the DSL. 

Introducing redundancy for validating the correctness of the generated artifacts requires extra effort to develop techniques that enable these checks and to analyze the results. However, these redundant mechanisms can effectively detect mistakes in the artifacts. Discovering mistakes through redundant mechanisms raises awareness about the potential inconsistencies between transformations from a DSL and induces search for other forms of redundancy to increase the reliability of the generated artifacts.

We have experimented with this approach as preparation for the redesign of a clinical X-ray generator. We plan to extract models from the existing implementation using model learning techniques. A learned model can give insight in the implemented behavior, and can be compared with the labeled transition system described by a DSL model.

\paragraph{Acknowledgement}
This research was supported by the Dutch national program
COMMIT and carried out as part of the Allegio project.

\bibliography{acquisition}

\appendix

\section{Validation Properties for \textit{Pedal Handling}}\label{app:properties}
In this appendix we provide some example properties that can be verified against specifications in the \textit{Pedal Handling} DSL. All the properties presented in this appendix are expressed in a variant of the modal $\mu$-calculus. This logic is the language used by the mCRL2 toolset \cite{mCRL2} for property specification. The interested reader can refer to \cite{GM14} for more details about the syntax and semantics of this logic.

We provide example properties for the DSL model of Fig.~\ref{fig:acq_dslexample}. As discussed in Section~\ref{sec:acq_casestudy}, a specification in the Pedal Handling DSL declares a set of input actions and describes the way the state of the system evolves as it perform the actions. For example the DSL model of Fig.~\ref{fig:acq_dslexample} declares the input actions $\textit{FRFluoOn}, \textit{FRFluoOff}, \textit{StartCond}$, and $\textit{ResetStartCond}$. To express properties for this DSL model, we first explain the behaviors represented by the declared actions:
\begin{itemize}
\item
\textit{FRFluoOn}: request to start capturing real-time images;
\item
\textit{FRFluoOff}: request to stop capturing real-time images;
\item
\textit{StartCond}: notification to prevent X-ray from starting;
\item
\textit{ResetStartCond}: notification to allow X-ray generation for new requests.
\end{itemize}
\textit{Pedal Handling} performs actions $\textit{output}(\xr,p)$ to send X-ray requests for $\xr\in\xray$ to $p\in\plane$.

In what follows, we formalize some example properties for the DSL model of Fig.~\ref{fig:acq_dslexample}. The formalized properties assume that there is a data specification defining the data types $\xray$ and $\plane$ (see Section~\ref{subsec:acq_domain}) and refer to the actions discussed above.

\begin{itemize}
\item
In every reachable state there is at least one outgoing transition (deadlock-freedom):
\begin{align*}
[\true^*]\langle \true \rangle \true
\end{align*}
\item
no X-ray is generated from the planes when there is no request from the user:
\begin{align*}
\nu X&(f:\bool=\textit{false}).(\\
& [\textit{FRFluoOn}]X(\true) ~\wedge\\
& [\textit{FRFluoOff}]X(\textit{false})~\wedge\\
& \bigg(\forall \xr:\textit{XRay},p:\plane . [{\it output}(\xr,p)]\Big(X(f)~\wedge \\
&\phantom{(\forall \xr:\textit{XRay},p:\plane . [{\it output}(}(\neg f \Rightarrow (\xr=\textit{Standby} \wedge p=\textit{None}))\Big)\bigg)~\wedge	 \\	
& [\overline{\textit{FRFluoOn}} ~\wedge ~ \overline{\textit{FRFluoOFF}} ~\wedge ~ (\forall \xr:\textit{XRay},p:\plane . \overline{{\it output}(\xr,p)})]X(f))
\end{align*}
\item
if there is a start condition (i.e., a \textit{StartCond} notification that is not canceled by a \textit{ResetStartCond}), X-ray generation cannot be started. It should be noted that a \textit{StartCond} notification does not interrupt the X-ray generation:
\begin{align*}
\nu X&({\it fr}:\bool=\false,{\it sc}:\bool=\false).(\\
&[{\it FRFluoOn}]X(\true,{\it sc})~\wedge \\
&[{\it FRFluoOff}]X(\false,{\it sc})~\wedge \\
&[{\it StartCond}]X(\false,\true)~\wedge\\
&[{\it ResetStartCond}]X(\false,\false)~\wedge \\
&\bigg(\forall \xr:\xray,p:\plane.[{\it output}(\xr,p)]\Big(X(\false,{\it sc})~\wedge\\
&\phantom{(\forall \xr:\xray,p:\plane.[{\it output}} \big(({\it fr}\wedge {\it sc})\Rightarrow (\xr={\it Standby} ~\wedge ~ c={\it None}))\Big)\bigg)
\end{align*}

\end{itemize}

\end{document}